\documentclass[onecolumn,draftclsnofoot,12pt]{IEEEtran}
\usepackage{ifpdf}
\usepackage{cite}
\usepackage[cmex10]{amsmath}
\usepackage{array}
\usepackage{amsfonts}
\usepackage{mathrsfs}
\usepackage{arydshln}
\usepackage{slashbox}
\usepackage{graphicx}
\usepackage{float}
\usepackage{subfigure}
\usepackage{amssymb}
\usepackage{amsmath}
\usepackage{url}
\usepackage{multirow}
\usepackage{multicol}
\usepackage{cite}
\usepackage[subfigure]{graphfig}
\usepackage{xcolor}
\usepackage{setspace}
\newcommand{\mv}[1]{\mbox{\boldmath{$ #1 $}}}
\usepackage[linesnumbered,ruled]{algorithm2e}

\newtheorem{lemma}{\underline{Lemma}}
\newtheorem{proposition}{\underline{Proposition}}[section]

\newcommand{\qed}{\nobreak \ifvmode \relax \else
      \ifdim\lastskip<1.5em \hskip-\lastskip
      \hskip1.5em plus0em minus0.5em \fi \nobreak
      \vrule height0.75em width0.5em depth0.25em\fi}

\begin{document}
\title{{Cellular-Enabled UAV Communication: A Connectivity-Constrained Trajectory Optimization Perspective}\thanks{This work will be presented in part at the IEEE International Conference on Communications (ICC), Kansas City, MO, USA, May 20-24, 2018 \cite{cellularUAV_arXiv}.}\thanks{The
authors are with the Department of Electrical and Computer Engineering, National University of Singapore (e-mails: \{elezhsh;elezeng;elezhang\}@nus.edu.sg).}
\author{\IEEEauthorblockN{Shuowen~Zhang, \emph{Member, IEEE}, Yong~Zeng, \emph{Member, IEEE}, and
Rui~Zhang, \emph{Fellow, IEEE}}\\\emph{(Invited Paper)}}}
\maketitle

\begin{abstract}
Integrating the unmanned aerial vehicles (UAVs) into the cellular network is envisioned to be a promising technology to significantly enhance the communication performance of both UAVs and existing terrestrial users. In this paper, we first provide an overview on the two main paradigms in cellular UAV communications, namely, \emph{cellular-enabled UAV communication} with UAVs as new aerial users served by the ground base stations (GBSs), and \emph{UAV-assisted cellular communication} with UAVs as new aerial communication platforms serving the terrestrial users. Then, we focus on the former paradigm and study a new UAV trajectory design problem subject to practical communication connectivity constraints with the GBSs. Specifically, we consider a cellular-connected UAV in the mission of flying from an initial location to a final location that are given, during which it needs to maintain reliable communication with the cellular network by associating with one of the available GBSs at each time instant that has the best line-of-sight (LoS) channel (or shortest distance) with it. We aim to minimize the UAV's mission completion time by optimizing its trajectory, subject to a quality-of-connectivity constraint of the GBS-UAV link specified by a minimum receive signal-to-noise ratio (SNR) target, which needs to be satisfied throughout its mission. To tackle this challenging non-convex optimization problem, we first propose an efficient method to verify its feasibility via checking the connectivity between two given vertices on an equivalent graph. Next, by examining the GBS-UAV association sequence over time, we obtain useful structural results on the optimal UAV trajectory, based on which two efficient methods are proposed to find high-quality approximate trajectory solutions by leveraging the techniques from graph theory and convex optimization. The proposed methods are analytically shown to be capable of achieving a flexible trade-off between complexity and performance, and yielding a solution that is arbitrarily close to the optimal solution in polynomial time. Numerical results further validate the effectiveness of our proposed designs against benchmark schemes. Finally, we make concluding remarks and point out some promising directions for future work.
\end{abstract}

\section{Introduction}\label{sec_intro}
Featured by high mobility and flexible deployment, \emph{unmanned aerial vehicles (UAVs)}, or drones, have become appealing solutions for a wide range of commercial and civilian applications over the past few years, including traffic control, cargo delivery, surveillance, aerial inspection, rescue and search, video streaming, precision agriculture, etc \cite{UAS,handbook}. Particularly, UAVs are found extremely useful to replace humans or manned aircrafts in missions that are dull (e.g., extended surveillance), dirty (e.g., pesticide spray), and dangerous (e.g., rescue and search after disaster) \cite{UAS}. Driven by the continuous cost reduction in UAV manufacturing, as well as the recent government efforts in devising UAV-related regulations in many countries \cite{US,Singapore}, the demand for UAVs is expected to further skyrocket in the near future. For example, the global UAV market was valued at 18.14 billion U.S. dollars in 2017, and is projected to reach 52.30 billion U.S. dollars by 2025 \cite{UAVmarket}. Moreover, according to \cite{UAVshipments}, the number of consumer and commercial UAV shipments is expected to reach 29 million and 805,000 by 2021, respectively.

The prosperous market and unique features of UAVs are envisioned to bring both challenges and opportunities to the cellular industry. On one hand, to realize the unprecedentedly large-scale deployment of UAVs in the future, it is of utmost importance to ensure that all UAVs can fly safely, which requires ultra-reliable and secure communication links between UAVs and their ground control stations (GCSs)/pilots. Moreover, high-rate UAV-ground communication links are also in need for various rate-demanding applications, e.g., when UAVs need to timely send their captured high-resolution photo/video to the remote users on the ground. A promising solution to meet the above UAV communication requirements is by integrating UAVs into the existing and future-generation cellular networks as new aerial users to be served by the ground base stations (GBSs). On the other hand, thanks to the continuous improvement in UAV payload weight and communication device miniaturization, it becomes more feasible for UAVs (such as drones, helikites, balloons) to carry communication equipments in the sky, to provide or enhance the communication services for the terrestrial users in the cellular networks (see e.g. \cite{FB,Google} for initial trials made by Facebook and Google). Generally speaking, the interplay between UAVs and the cellular network can be fully exploited to improve the communication performance of both UAVs and existing cellular users on the ground, which calls for a new research paradigm to investigate their joint design and performance optimization.

In the rest of this section, we first give an overview on the promising technologies for {\hbox{integrating}} UAVs into the cellular network. Next, we provide a review of related prior work, and discuss a new and critical communication-aware trajectory design problem in cellular-enabled UAV communication, which motivates the main research in this work. Last, we {\hbox{summarize}} the main results and contributions of this paper for solving this problem.

\subsection{UAV Meets Cellular Network: An Overview}
As briefly introduced above, there are two main paradigms for integrating UAVs into the cellular network, depending on the role that UAVs play:
\begin{itemize}
\item {\bf{Cellular-enabled UAV communication}}: UAVs with their own missions (e.g., cargo delivery, video surveillance) operate as new \emph{aerial users}, which are served by the cellular GBSs.
\item {\bf{UAV-assisted cellular communication}}: UAVs operate as new \emph{aerial communication platforms} (e.g., aerial base stations (BSs), relays) to serve the terrestrial users in the cellular network.
\end{itemize}
In this subsection, we provide a detailed overview on the motivations and benefits for these two types of cellular UAV communications, as well as the challenging issues that need to be addressed for UAV and cellular network designs, respectively.

\subsubsection{Cellular-Enabled UAV Communication}
Ensuring safe and efficient operation of UAVs requires ultra-reliable, low-latency, and highly secure communication links between the UAVs and their GCSs, in order to support their two-way \emph{control and non-payload communication (CNPC)} \cite{ITU,CNPC}. Specifically, CNPC consists of
the following types of information flows between the UAV and GCSs, which are crucial to the UAV operation: command and control data (e.g., uplink telecommand and downlink telemetry data, uplink navigation setting changes and downlink navigation display data), air traffic control (ATC) relay
data (e.g., uplink ATC voice and air traffic service (ATS) messages), sense and avoid (S$\&$A) data (e.g., downlink S$\&$A target track data, airborne weather radar data, video for improved pilot awareness in special situations such as take-off and landing), etc \cite{ITU}. On the other hand, in many practical applications such as video streaming, aerial surveillance, and data backhaul, UAVs generally require high-capacity data communication links with the ground terminals so as to timely
transmit the payload data (such as high-quality video and high-rate backhauled data). In contrast to CNPC with relatively low rate requirement, such mission-specific \emph{payload communication (PC)} typically demands for much higher transmission rate.

However, at present, almost all UAVs in the market rely on the simple direct point-to-point communication links with their ground pilots/GCSs over the unlicensed spectrum, e.g., ISM (Industrial, Scientific, and Medical) 2.4 GHz, for both CNPC and PC. Such links are typically of limited data rate and low reliability, vulnerable to interference, insecure, and can only operate within the visual line-of-sight (VLoS) range. Moreover, legitimate monitoring and management of these links are practically difficult. It was shown by recent studies that the dominant reason for the 150 reported UAV crashes over the past ten years is the loss of UAV-ground communication \cite{incidents}, which further affirms the limitation of the current approach. As the number of UAVs and the demand for new UAV applications are expected to increase explosively in the near future, it is of vital importance to devise innovative solutions to support high-performance communications between UAVs and the ground.

\begin{figure}[t]
  \centering
  \includegraphics[width=15cm]{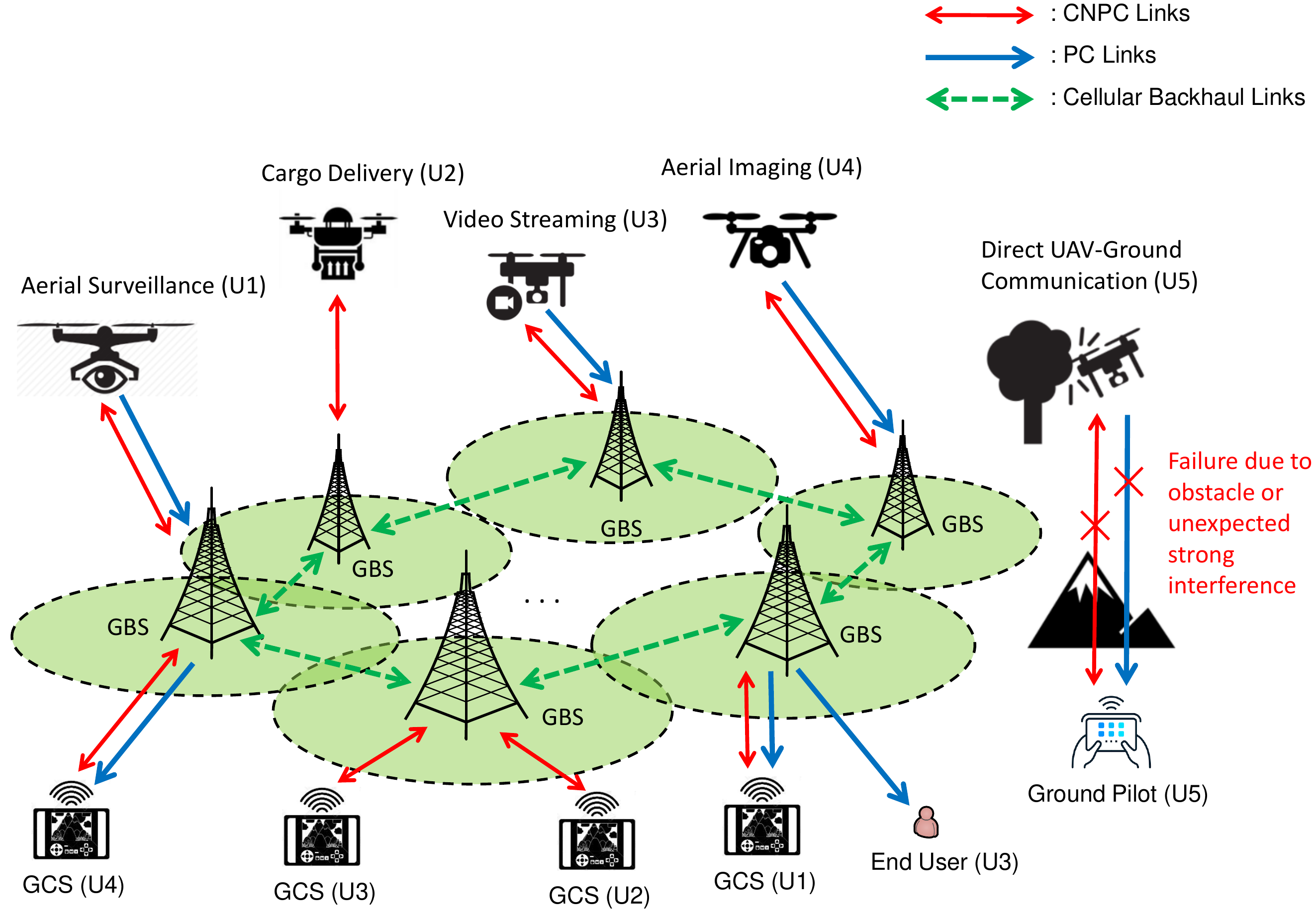}
  \caption{Schematic diagram for cellular-enabled UAV communication (where UAVs serve as aerial users) with four typical use cases and comparison with direct UAV-ground communication.}\label{Cellular_Enabled}
\end{figure}

To achieve the above goal, a promising approach is \emph{cellular-enabled UAV communication}, as illustrated in Fig. \ref{Cellular_Enabled}, where GBSs in the existing 4G (fourth-generation) LTE (Long Term Evolution) or the forthcoming 5G (fifth-generation) and beyond 5G (B5G) cellular networks are utilized to support communications between UAVs and their corresponding GCSs/end users. Thanks to the almost ubiquitous accessibility worldwide and superior performance of today's LTE and future 5G/B5G wireless networks, cellular-enabled UAV communication is expected to achieve significantly enhanced performance over the existing point-to-point UAV-ground communications, in terms of various performance metrics such as reliability, security, coverage and throughput. In particular, it potentially enables the safe and reliable CNPC links with flying UAVs for beyond VLoS (BVLoS) operations, which significantly extends the UAV operation range. Moreover, with the high-capacity and ubiquitous cellular connectivity, UAVs are able to communicate with potentially gigabits per second (Gbps) link speed to end users that are even thousands of kilometers away, thus opening up many new applications for UAVs in the future. In addition, the advanced authentication mechanisms in cellular networks can be leveraged to achieve more effective and secure air traffic management. Last but not least, this solution is economically appealing since incorporating UAVs as new aerial users in LTE/5G/B5G brings new business opportunities to the wireless industry, for both device manufacturers and cellular service providers. The feasibility of supporting UAVs using the LTE network has already been demonstrated via preliminary measurement results in industry and academia \cite{Qualcomm,LTEsky,LinSky,CellularConnected}. 3GPP (the 3rd Generation Partnership Project) has recently approved the work item on enhanced support for aerial vehicles using LTE \cite{3GPPLTE}.

Despite its promising future, many new design challenges need to be tackled for realizing effective cellular-enabled UAV communication, due to the considerably different characteristics between the new aerial users and the conventional terrestrial users. Specifically, the most notable features of UAV-ground communication include:
\begin{itemize}
  \item \emph{High altitude}: UAVs typically operate at high altitude ranging from several meters to a few kilometers \cite{UAS}, which are in general much higher and vary much more drastically compared to the ground users.
  \item \emph{Line-of-sight (LoS)-dominant channel}: Different from conventional terrestrial communication channels that usually suffer from more severe attenuation over distance, shadowing and fading due to multi-path scattering, the UAV-GBS channels are typically LoS-dominant due to the high altitude of UAVs, especially in rural and suburban environments \cite{Qualcomm,RadioChannel,ASuburban}. Moreover, in urban environments, the probability of having LoS links generally increases with the UAV altitude \cite{ModelingA2G,LAP,RadioChannel,ASuburban}.
  \item \emph{High and controllable mobility in three-dimensional (3D) airspace}: In contrast to ground users, UAVs possess the unique ability of flexibly moving in the 3D space with high speed and controllable trajectory.
  \item \emph{Asymmetric uplink and downlink communication requirements}: Different from the current cellular network, which is mainly designed to meet the dominant downlink data demand for downloading applications, cellular-enabled UAV communication needs to support higher data rate in the uplink (i.e., from UAVs to GBSs) for the UAVs to upload their mission-related data to the GBSs. On the other hand, the downlink (i.e., from GBSs to UAVs) CNPC link typically has much lower rate demand than the uplink PC, but more stringent delay and reliability requirement in general.
\end{itemize}
Stemming from these features, the following new design challenges arise:
\begin{itemize}
\item \emph{3D coverage}: In the current 4G LTE cellular network, GBS antennas are generally tilted downwards, in order to provide satisfactory \emph{2D coverage} to the ground users with large antenna gain and reduced inter-cell interference. However, as UAVs may fly at higher altitude than the GBSs, they may need to be served by the GBS antenna side-lobes or reflected signals when connected to the LTE network, which results in significantly weaker antenna gain as compared to the ground users. Moreover, in certain applications, UAVs may need to fly over remote rural areas or even unpopulated areas, where only very limited or no cellular coverage exists. How to achieve seamless and high-quality \emph{3D coverage} for both aerial and ground users is thus a challenging problem, which calls for new antenna design solutions and revised GBS deployment strategies for the future 5G and B5G cellular networks, e.g., one possible approach to improve the 3D coverage is to deploy dedicated GBS antennas that are tilted upwards for serving the aerial users.
\item \emph{Aerial interference management}: The high UAV altitude and LoS-dominant channels between GBSs and UAVs are expected to result in more severe interference in cellular networks as compared to that caused by conventional terrestrial communications only. As verified by some preliminary measurement results, the number of GBSs that a UAV may detect in an LTE network increases with the UAV altitude \cite{Qualcomm,LTEsky}. These include not only the intended GBSs that can be associated with the UAV for communication (hence, yielding a higher macro-diversity gain compared to ground users), but also non-associated GBSs with potentially strong interference. Consequently, in the downlink transmission, the UAV may receive severe interference from many non-associated GBSs; while on the other hand, in the uplink transmission, the UAV may cause strong interference to the non-associated GBSs as well. This severe aerial-ground interference issue, if not tackled properly, could significantly limit the performance of UAV communication and even degrade dramatically the performance of existing terrestrial communication, thus leading to an overall spectral efficiency loss for the cellular network. To overcome this issue, new and effective interference management techniques need to be developed.
\item \emph{Communication-aware trajectory design}: Compared to ground users, the 3D mobility of UAVs offers a new design degree-of-freedom (DoF) for improving the communication performance. Specifically, the UAV trajectory can be designed to meet its mission specifications (e.g., travelling between a pair of locations before a specified deadline), while at the same time ensuring the communication requirements for CNPC and/or PC during its mission. This gives rise to a new approach of communication-aware trajectory design for UAVs, which can be used to solve the aforementioned challenges. For example, the UAV trajectory can be flexibly designed based on the known locations of the GBSs in its fly direction as well as the distribution of the ground users to ensure its communication coverage by the associated GBSs and at the same time reduce the interference to/from the ground users/non-associated GBSs.
\end{itemize}

\subsubsection{UAV-Assisted Cellular Communication}
Besides enhancing the communication performance of UAVs by leveraging cellular-enabled UAV communication, integration of UAVs into the cellular network can also be exploited to enhance the quality-of-service (QoS) of the existing terrestrial cellular users, by deploying UAVs as new aerial communication platforms (e.g., aerial BSs and relays) to assist in the terrestrial cellular communications \cite{survey}. Compared to the existing cellular communication infrastructures (e.g., GBSs, fixed relays), UAV-mounted aerial communication platforms possess the following new features:
\begin{itemize}
\item \emph{Swift and flexible deployment}: Conventional terrestrial communication infrastructures are generally stationary, i.e., their locations are fixed once they are deployed, thus they can be designed only based on the long-term data traffic and user distribution. On the contrary, UAVs can be flexibly deployed as \emph{quasi-stationary aerial BSs/relays} whose locations can be dynamically adjusted according to the real-time demand, which enables swift network reconfiguration based on the temporal data traffic/user locations and consequently leads to enhanced performance. Particularly, the high flexibility and on-demand deployment of UAVs make them an attractive solution to provide ubiquitous cellular coverage for remote areas or temporary events, without the need of building new terrestrial communication infrastructure. Moreover, in contrast to fixed GBSs/relays whose deployments are restricted to the 2D ground plane under terrain-specified constraints, UAVs can be more flexibly deployed in the 3D airspace with adjustable altitude.
\item \emph{High and controllable 3D mobility}: Besides serving as quasi-stationary aerial BSs/relays at given locations, the high and controllable 3D mobility of UAVs enable them to serve as \emph{mobile aerial BSs/relays} that are capable of flying contiguously over the serving area to communicate with the ground users more efficiently. In this case, the UAV trajectory design is generally more flexible as compared to that in cellular-enabled UAV communication where UAVs usually have mission-specific constraints on their trajectories.
\item \emph{LoS-dominant channel}: Similar to the GBS-UAV channels in cellular-enabled UAV communication, the channels between UAV communication platforms and ground users in UAV-assisted cellular communication are also dominated by the LoS links in rural and suburban environments, or can be characterized via a probabilistic LoS model in an urban environment \cite{ModelingA2G,LAP}. The absence of fading generally leads to more stable link quality and improved reliability as compared to conventional terrestrial communication in rich-scattering environments.
\end{itemize}

\begin{figure}[t]
  \centering
  \includegraphics[width=17cm]{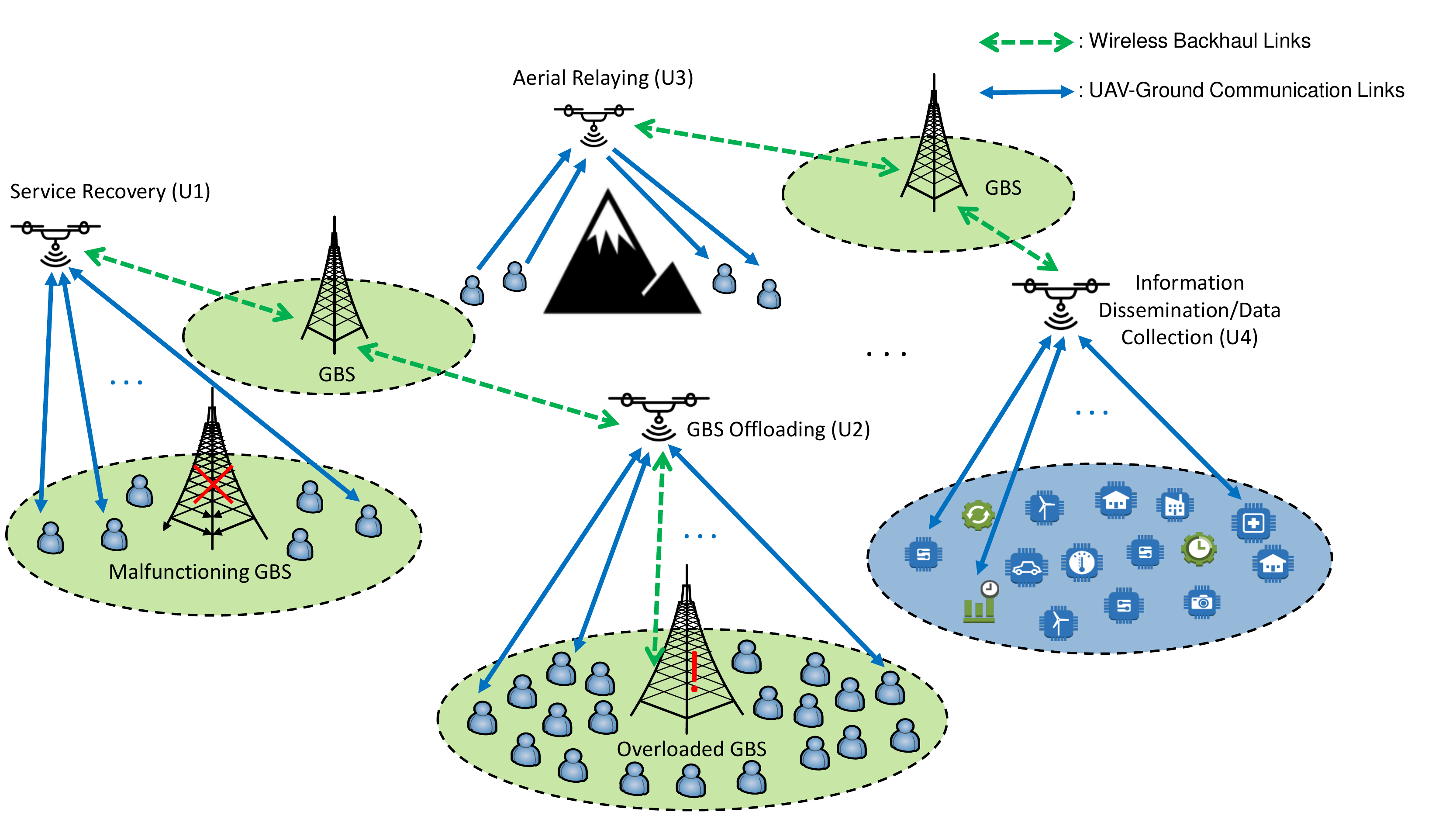}
  \caption{Schematic diagram for UAV-assisted cellular communication (where UAVs
  serve as aerial communication platforms) with four typical use cases.}\label{UAV_Assisted}
\end{figure}

Due to the above advantages, UAV-assisted cellular communication is envisioned to be an important technology for future cellular networks to satisfy the more dynamic and diversified communication demands. Some typical use cases include: cellular coverage expansion in remote areas without ground communication infrastructure; swift service recovery when the ground communication infrastructure is malfunctioning due to natural disasters in emergency situations; GBS data traffic offloading in temporary hotspot areas; aerial relaying between distant ground user clusters; and cost-effective information dissemination and data collection for Internet-of-Things (IoT) and machine-type communications (MTC), etc \cite{survey}, which are illustrated in {\hbox{Fig. \ref{UAV_Assisted}.}}

To realize the promising vision of UAV-assisted cellular communication, the following new design challenges need to be addressed, which are usually overlooked in conventional terrestrial communications:
\begin{itemize}
\item \emph{3D deployment for quasi-stationary UAVs}: The 3D deployment for quasi-stationary UAV communication platforms is more challenging than the conventional 2D placement of {\hbox{GBSs}}/ relays, due to the additional design DoF in the UAV altitude, as well as the more flexible horizontal deployment of the UAV in the absence of ground obstacles. Moreover, the dynamic UAV deployment to cater for the change of ground network topology is also an important design problem. In general, due to the lack of wired backhauls which are available for the GBSs/relays, UAV communication platforms need to rely on wireless backhauls to inter-connect as well as connect to the GBSs/gateways (see Fig. 2). As a result, the UAV deployment problem should also take into account the wireless backhaul capacities, which are usually time-varying due to the mobility of UAVs.
\item \emph{Communication-oriented trajectory design for mobile UAVs}: The trajectory of mobile UAV communication platforms needs to be properly designed to maximize the communication performance of the ground users. For example, a UAV can fly closer to a ground user with high rate requirement to shorten their communication link distance, so as to increase the link capacity or save the transmit power. Moreover, the UAV trajectory can be jointly designed with communication resources allocation to further improve the performance. In this regard, the severe interference issue among multiple serving UAVs as well as with the existing GBSs due to their LoS channels needs to be resolved.
\item \emph{Energy-efficient communication considering UAV's propulsion energy consumption}: Besides the conventional communication-related energy consumption due to signal processing and amplification, UAV communication platforms are subjected to the additional propulsion energy consumption for them to remain aloft and move freely, which is usually much more significant than the communication energy (e.g., in the order of kilowatt versus watt) \cite{energy,Rotary}. Unlike ground communication infrastructures which usually have abundant power supply, the limited on-board energy due to the stringent SWAP (size, weight, and power) constraint and high propulsion energy consumption of UAVs pose critical limits on their communication performance and endurance. Thus, energy-efficient design is crucial for enhancing the long-term performance of UAV-assisted cellular communication.
\end{itemize}

\subsection{Literature Review}
Motivated by the appealing applications and new design challenges of cellular UAV communications as discussed in the preceding subsection, there has been an upsurge of interests in the research on both paradigms over recent years, for which we provide a survey on the related prior work, respectively, as follows.
\subsubsection{Cellular-Enabled UAV Communication}
Despite its high potential and intensive industrial interests, research in this area is still in an infancy stage. The feasibility of supporting aerial users with 4G LTE networks was studied in \cite{LTEsky,LinSky,CellularConnected}, where initial performance evaluation was carried out via measurements and simulations. It was shown in \cite{LTEsky} that for a network with coexisting aerial and ground users, the performance of both types of users generally degrades as the aerial user altitude increases due to the more severe aerial interference. Moreover, it was shown in \cite{LinSky,CellularConnected} that the increase in the number of aerial users leads to stronger aerial interference and consequently decreases the network spectral efficiency. These results demonstrated that aerial interference is one major performance limiting factor in LTE-enabled UAV communication with the existing antenna patterns at the GBSs, which are typically fixed and tilted downwards to cover the terrestrial users. The effect of aerial interference was also investigated analytically in \cite{PCoexistence}, where the coverage probability for both aerial users and ground users coexisting in an LTE downlink was characterized. It was shown in \cite{PCoexistence} that reducing the UAV flying altitude, lowering the GBS heights, and increasing the GBS antenna downtilt angle are beneficial for aerial interference mitigation and consequently the network performance. The results in \cite{PCoexistence} were extended in \cite{PReshaping} by considering a different GBS association rule based on the channel strength between GBS and UAV instead of their geographical distance as considered in \cite{PCoexistence}. Initial attempts for aerial interference mitigation were made in \cite{Ensure,CellularConnected}, where the potential of several techniques including 3D beamforming, antenna beam selection, interference cancellation, power control, and inter-cell interference coordination was validated by simulations. Furthermore, an interference-aware path planning design was proposed in \cite{MLearning} for a cellular-enabled multi-UAV communication system, which aimed to achieve a trade-off between energy efficiency, latency, and interference caused by the UAV to the ground network. In \cite{LSwarm,LMassive}, massive multiple-input multiple-output (MIMO) was studied as a promising solution to enhance the 3D coverage range and support the simultaneous operation of multiple UAVs, by employing a large size of arrays at the GBSs. Specifically, the uplink capacity of a massive MIMO GBS enabled multi-UAV communication system was derived in \cite{LSwarm}, based on which the optimal antenna spacing at the GBS was determined.

\subsubsection{UAV-Assisted Cellular Communication}
In contrast to cellular-enabled UAV communication, UAV-assisted terrestrial/cellular communication has received research attentions since much earlier and there have been rich literatures in this paradigm, for which a detailed survey is given as follows.

On one hand, \emph{3D deployment of UAV-mounted quasi-stationary aerial communication platforms} has been studied in \cite{Flocking,HanMANET,LAP,PJoint,POptimal,H3D_ICC16,HBackhaul,H3D_EE,H3D,placement,HNumber,MSDeployment,AltitudeBeamwidth,CoMP}. In \cite{Flocking}, a local flocking based strategy was proposed to optimize the UAV placement and navigation for maximizing its connectivity with the ground users. A mobile ad hoc network (MANET) with a flying UAV was considered in \cite{HanMANET}, where the UAV deployment and movement were designed to optimize the ground network connectivity with various measures. In \cite{LAP}, the optimal UAV altitude that maximizes its ground coverage area was derived under a probabilistic LoS channel model. Under the Rician fading channel model, the optimal UAV altitude for rate maximization or outage probability minimization was derived in \cite{PJoint,POptimal}. Under the probabilistic LoS channel model, the 3D deployment of a single UAV was investigated in \cite{H3D_ICC16}, in order to maximize the number of covered users in a given area. The work in \cite{H3D_ICC16} was extended in \cite{HBackhaul,H3D_EE,H3D}, with additional considerations of the backhaul rate and bandwidth constraint for the UAV \cite{HBackhaul}, transmit power consumption \cite{H3D_EE}, or heterogeneous QoS constraints for the users \cite{H3D}. Moreover, the joint deployment of multiple UAVs was studied in \cite{placement}, aiming to minimize the number of required UAVs while ensuring that all ground users are covered, under the LoS channel model. An efficient spiral-based algorithm with polynomial complexity was proposed in \cite{placement}, which was shown to achieve close performance to the optimal solution requiring exponential complexity. The same problem was also considered in \cite{HNumber} under the probabilistic LoS channel model, where a heuristic algorithm based on particle swarm optimization was proposed. In \cite{MSDeployment}, the 3D deployment of multiple UAVs to maximize the total ground coverage area was studied, under the assumption that a directional antenna with fixed beamwidth is equipped at the UAV instead of an omnidirectional antenna as considered in \cite{Flocking,HanMANET,LAP,PJoint,POptimal,H3D_ICC16,HBackhaul,H3D_EE,H3D,placement,HNumber}. Furthermore, \cite{AltitudeBeamwidth} considered the scenario where a UAV with adjustable altitude and tunable antenna beamwidth is deployed to serve multiple ground users under a proposed ``fly-hover-and-communicate'' protocol. The joint altitude and beamwidth optimization was studied under three fundamental multiuser communication models, namely, downlink multicasting, downlink broadcasting, and uplink multiple access. It is worth noting that all the above work considered a single UAV or multiple UAVs without cooperation. In \cite{CoMP}, a novel architecture termed ``coordinated multipoint (CoMP) in the sky'' was proposed, where multiple UAVs are deployed to cooperatively serve multiple ground users with joint transmission/reception. A realistic LoS channel model with random phase was proposed in \cite{CoMP}, under which the placement and movement of multiple UAVs were jointly optimized to maximize the network throughput.

On the other hand, \emph{communication-oriented trajectory optimization for UAV-mounted mobile relays} has been investigated in \cite{RelayVlah,Airborne,RelayBrown,RelayGuo,RelayALS,RelayTsourdos,relay,Song,DAlgorithm,DSegmented,DPositioning}. In \cite{RelayVlah}, a UAV-enabled mobile relay system was considered under a ``load-carry-and-deliver'' strategy, where the UAV trajectory was optimized to maximize the throughput under different data delay requirements. In \cite{Airborne,RelayBrown}, the deployment of a UAV for ferrying data from multiple sources to a destination was considered, where a policy gradient reinforcement learning algorithm for the UAV trajectory design was proposed towards delay minimization. In \cite{RelayGuo}, a two-hop relay multi-session routing scheme was proposed, where a moving vehicle serves as a relay to deliver independent data packets to multiple user groups. The relay trajectory in terms of its sequence of visiting the user groups was jointly optimized with the packet size to maximize the worst performance among all groups. In \cite{RelayALS}, multiple UAVs were deployed as relays between multiple ground terminals and a network BS. Closed-form solutions of the UAV headings that maximize the sum network data rate while ensuring a satisfactory rate requirement of each individual relay link were derived. In \cite{RelayTsourdos}, coordinated trajectory planning for multiple UAVs as mobile relays in a MANET was studied under various UAV operational and dynamic constraints to optimize the network connectivity. Note that in the above work, the relay trajectory was generally not jointly optimized with the communication resource allocation for performance optimization. An initial attempt along this line was pursued in \cite{relay}, where a classic three-node relay system with the UAV serving as a mobile decode-and-forward (DF) relay was considered. A rigorous mathematical problem was formulated to jointly optimize the UAV trajectory and source/relay transmit power. As the formulated problem is non-convex, the block coordinate descent method was applied to update the transmit power and UAV trajectory iteratively, and the successive convex approximation (SCA) technique was leveraged to find high-quality approximate solutions for the trajectory optimization sub-problem. As an extension to \cite{relay}, a UAV-mounted mobile amplify-and-forward (AF) relay was considered in \cite{Song}, where the UAV trajectory and transmit power were jointly optimized to minimize the outage probability of the relay network. While offline trajectory design was considered in the above work under the LoS channel model, trajectory of UAV-mounted mobile relays can also be autonomously designed online based on real-time measurements of the UAV-ground channels \cite{DAlgorithm,DSegmented,DPositioning}, which is especially suitable for dense urban environments where the UAV-ground channels may experience shadowing and fading.

Furthermore, \emph{trajectory and communication resource allocation optimization for UAV-mounted mobile BSs} has been considered in \cite{cyclical,Aviation,multicast,Zhan,JointTC,delay,Capacity}. In \cite{cyclical,Aviation}, a UAV-enabled multiuser communication system was considered, where a UAV-BS is deployed to serve multiple ground users located on a straight line by following a straight flight trajectory. A cyclical time-division multiple access scheme was proposed in \cite{cyclical}, based on which a fundamental trade-off between throughput and access delay was revealed. Moreover, the total UAV aviation time to collect the data transmitted from the ground users was minimized in \cite{Aviation} by jointly optimizing the data collection interval and transmit power of the users, as well as the UAV speed. A UAV-enabled multicasting system was studied in \cite{multicast}, where a UAV-BS is deployed to disseminate a common file to multiple ground users. The UAV trajectory was optimized to minimize the mission completion time, while ensuring that the file can be recovered at every user with a satisfactorily high probability. The concept of virtual BS placement and techniques for solving the traveling salesman problem with neighbourhood (TSPN) were utilized to find efficient UAV trajectories. In \cite{Zhan}, a UAV-enabled wireless sensor network (WSN) was considered, where a UAV-BS is dispatched to fly over multiple sensor nodes for estimating an unknown parameter based on its collected data. An efficient suboptimal UAV trajectory design for optimizing the estimation performance was proposed. In \cite{JointTC,delay,Capacity}, the UAV trajectory and communication resource allocation were jointly designed. Specifically, in \cite{JointTC}, a multi-UAV enabled multiuser communication system was considered, where the UAV trajectory and power control, as well as user scheduling and association were jointly optimized to maximize the minimum throughput over all users for the downlink transmission scenario. Moreover, a UAV-enabled multiuser orthogonal frequency division multiple access (OFDMA) system with user delay consideration was studied in \cite{delay}, where the network throughput gain by exploiting the UAV mobility was shown diminishing as the user delay requirement becomes more stringent. In \cite{Capacity}, a UAV-enabled two-user broadcast channel was considered, and an information-theoretic approach was taken to characterize the capacity region of this simpled multiuser system. It was shown that a practical and low-complexity ``hover-fly-hover'' trajectory achieves the capacity.

It is worth noting that as a related line of research, UAV path planning for dedicated missions without communication performance consideration has been studied in \cite{MIP,HowCollision,EnergyAware,MotionPlanning}, where the UAV paths were designed such that the UAVs can reach their pre-determined destinations while avoiding collisions with other UAVs and/or terrain obstacles \cite{MIP,HowCollision}, or ensuring that a given set of points can be covered by their paths \cite{EnergyAware}. Moreover, there is another related line of research on the connectivity-seeking path planning for mobile robots \cite{First,YConnectivity}, where the ground robot communication channels possess different characteristics compared to the UAV-ground channels in cellular UAV communications; thus the UAV trajectory design problem formulations and solutions under the two distinct channel models are also different in general.

It is also worth noting that there is another line of work on the design and analysis of \emph{cellular networks with coexisting UAV-mounted BSs/relays and conventional GBSs} \cite{HDroneCell,MultiTier,offloading,WeiJSAC,HAssociation,HStrategic,MSTransportCell}, among which general overviews can be found in \cite{HDroneCell,MultiTier}. In \cite{offloading}, a new hybrid network architecture was proposed to resolve the cellular hotspot issue, where a UAV flies along the GBS cell edge for data offloading. The spectrum allocation, user partitioning, and UAV trajectory were jointly designed to maximize the common throughput of all users, which was shown to significantly outperform the conventional offloading schemes based on static small-cells. In \cite{WeiJSAC}, the coverage probability and achievable throughput of a drone small cell (DSC) network were characterized for both cases with and without GBSs, based on which the optimal DSC density for throughput maximization was derived. In \cite{HAssociation}, coexistence of a macro GBS (MBS) and several UAV-BSs that rely on wireless backhaul links with the MBS was studied, where the 3D UAV-BS deployment, user-BS association, and wireless backhaul bandwidth allocation were jointly optimized to maximize the user rate. A strategic placement of UAV-BSs over existing network of GBSs was proposed in \cite{HStrategic} based on a stochastic geometry approach. In \cite{MSTransportCell}, the cell association optimization among UAV-BSs and GBSs to minimize the average network delay was investigated, for which the optimal solution was proposed by leveraging the optimal transport theory.

Besides, UAV-assisted cellular communication with the \emph{UAV propulsion energy consumption consideration} has been investigated in \cite{energy,Rotary,spectrum,MEC}. In \cite{energy}, a mathematical model for the propulsion energy consumption of fixed-wing UAVs was derived as a function of the UAV velocity and acceleration. Based on the proposed model, trajectory optimization of a single UAV serving one ground user was studied in \cite{energy} to maximize the energy efficiency (EE), namely, the total transmitted information bits normalized by the energy consumption in Joule. This work has been extended to rotary-wing UAVs in \cite{Rotary}, where a new energy model was derived, and the UAV energy consumption was minimized while ensuring the users' communication throughput requirements. In \cite{spectrum}, the UAV trajectory and receiving/forwarding time allocation in a UAV-enabled mobile relaying system were jointly optimized for spectrum efficiency (SE) maximization and EE maximization, respectively, where a fundamental trade-off between SE and EE maximization was revealed. In \cite{MEC}, a UAV-mounted cloudlet was considered in a mobile edge computing system for providing offloading service to multiple ground users, where the UAV trajectory and uplink/downlink bit allocation were jointly optimized to minimize the total mobile energy consumption, including the computation, communication, as well as flying energy consumption. Moreover, energy consumption at the ground users served by the UAV was considered in \cite{MSIoT,sensorZhan,EnergyTradeoff}. In \cite{MSIoT}, the placement and movement of multiple UAVs collecting data from multiple IoT devices were jointly optimized with the device association and power control to minimize the total transmit power of the IoT devices. In \cite{sensorZhan}, a WSN with a UAV serving as a mobile data collector was considered, where the maximum energy consumption among all the sensor nodes was minimized via jointly optimizing the sensor nodes' wake-up schedule and the UAV trajectory. A UAV-enabled data collection system with one pair of UAV and ground node was considered in \cite{EnergyTradeoff}, where the UAV trajectory and ground node transmit power were jointly optimized to reveal a new fundamental trade-off between the UAV propulsion energy consumption and ground node communication energy consumption.

Finally, it is worth noting that UAV-assisted cellular communication can be integrated with {\hbox{other}} promising technologies for meeting the heterogeneous service demands in different application scenarios \cite{XiammWave,MSD2D,UAVWPT,WPCN,MSCaching,overcome,SecuringUAV,wiretap,GUAV,GImproved,GLTE,GTransport}. In \cite{XiammWave}, \emph{millimeter wave (mmWave)} was proposed as a suitable candidate for realizing high-speed UAV-assisted cellular communication, by exploiting the favorable LoS propagation condition. In \cite{MSD2D}, the coexistence of a UAV-mounted aerial BS and an underlaid \emph{device-to-device (D2D)} network was considered, where the coverage probability of the ground users was derived and analyzed. In \cite{UAVWPT}, a UAV-enabled \emph{wireless power transfer (WPT)} system was considered, where the trajectory of a UAV-mounted energy transmitter was optimized for maximizing the total amount of energy transmitted to the ground energy receivers. A UAV-enabled \emph{wireless powered communication network (WPCN)} was studied in \cite{WPCN}, where a UAV is deployed to transfer wireless energy to multiple ground users in the downlink, and receive information in the uplink sent by the users using their received energy. The \emph{proactive caching} technique was considered in \cite{MSCaching,overcome} for UAV-enabled communication, where the proactive deployment of cache-enabled UAVs was studied in \cite{MSCaching} to optimize the quality-of-experience (QoE) of ground users, while caching at the ground users was leveraged in \cite{overcome} to overcome the UAV endurance issue. For \emph{secrecy communications} from a physical-layer design perspective, a system with one UAV transmitter and a pair of ground legitimate receiver and ground eavesdropper was studied in \cite{SecuringUAV}, where the joint UAV trajectory and transmit power optimization was shown to effectively improve the secrecy rate; moreover, a UAV-enabled mobile jamming system was proposed in \cite{wiretap} to improve the secrecy rate of a ground wiretap channel. Last, UAV-assisted cellular communication was also leveraged in public safety communications \cite{GUAV,GImproved,GLTE} and intelligent transportation systems for smart city applications \cite{GTransport}.

Note that despite the rich prior work on cellular-UAV integration as reviewed above, communication-aware UAV trajectory design in cellular-connected UAVs is a relatively new problem, for which there is only limited work available in the literature \cite{cellularUAV_arXiv,disconnectivity,MLearning}. Notice that the trajectory design for UAV as a user is generally different from that with UAV as a BS/relay due to their different objectives: to be optimally served by the GBSs versus optimally serving the terrestrial users, along its trajectory. In general, the problems for designing the UAV trajectory to optimize a variety of communication performance metrics (such as throughput, delay, outage, energy efficiency) for cellular-connected UAVs subject to practical UAV energy and mobility constraints (including those on the endurance, altitude, speed, acceleration, etc.) still remain largely open, which motivates our main study in this work.

\subsection{Main Contributions}
\begin{figure}[t]
  \centering
  \includegraphics[width=13cm]{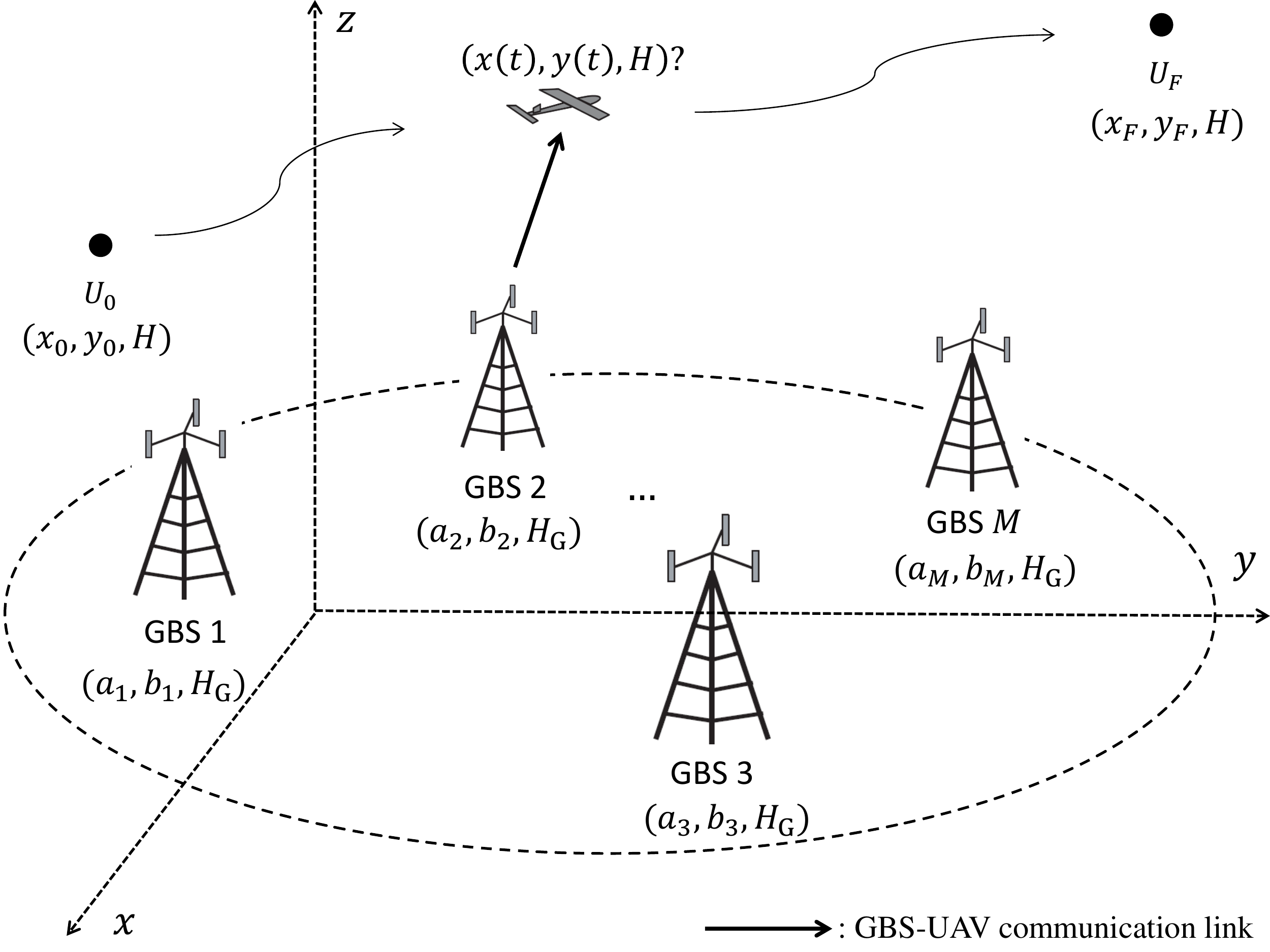}
  \caption{Illustration of a cellular-connected UAV with communication-aware trajectory design.}\label{cellularUAV}
\end{figure}
In this paper, we investigate a new and important UAV trajectory design problem in a cellular-enabled UAV communication system with a single UAV user communicating with the cellular network through the GBSs along its trajectory, as illustrated in Fig. \ref{cellularUAV}. The UAV has a mission of flying from an initial location to a final location that are assumed given, where it needs to achieve a target communication performance along its trajectory by associating with one GBS at each time instant which has the best channel with it among all available GBSs in its signal coverage area on the ground. Our main contributions are summarized as follows:
\begin{itemize}
\item We consider delay-sensitive communication between the UAV and the cellular network under a \emph{quality-of-connectivity} constraint specified by a minimum received signal-to-noise ratio (SNR) requirement which needs to be satisfied at all time along the UAV trajectory. This may correspond to the downlink command and control (C$\&$C) signal transmission from the GBSs to the UAV or the uplink payload data (e.g., real-time captured photo/video) transmission from the UAV to the GBSs. For simplicity, we assume that a dedicated frequency band is assigned for the exclusive use by this UAV, such that there is no interference to/from the GBSs that are not communicating with the UAV. Note that in practice the assigned band for the considered UAV can be opportunistically reused by the ground users and/or other UAVs sufficiently far away from this UAV such that their mutual interference can be safely neglected. Under the above setup, we formulate a UAV trajectory optimization problem to minimize the UAV mission completion time, subject to a given SNR requirement, the UAV's initial and final locations as well as its maximum speed. The formulated problem is non-convex and involves an infinite number of optimization variables, thus being difficult to solve in general.
\item First, we provide an efficient method for checking the feasibility of the formulated problem. Specifically, by examining the sequential GBS-UAV associations during the UAV mission, we show that the problem feasibility is equivalent to the connectivity between two {\hbox{given}} {\hbox{vertices}} in a properly constructed graph, which can be checked efficiently by applying existing algorithms in graph theory.
\item Next, with the feasibility of the formulated problem verified, we show that its optimal trajectory solution is achieved when the UAV flies by following a path consisting of connected line segments only with its maximum speed, where the UAV is associated with the GBS of the best channel with it while it flies within each line segment. By leveraging this optimal structure, the formulated problem is transformed into a joint optimization problem of the GBS-UAV association sequence and the UAV handover locations that specify the starting/ending points of each line segment along its trajectory. Compared to the originally formulated problem with infinite number of variables, this problem has a finite number of variables and is thus more tractable. Then, by applying appropriate bounding and approximation techniques, we introduce two graph construction methods, based on which the problem is further simplified to an equivalent \emph{shortest path problem} between two given vertices in the corresponding graph. Accordingly, two efficient methods are proposed for finding high-quality approximate solutions to the original problem by applying existing algorithms in graph theory and convex optimization techniques. It is shown analytically that our proposed methods are able to approach arbitrarily close to the optimal solution with moderately increasing complexity that is in a polynomial order of the number of available GBSs.
\item Moreover, numerical results are provided to validate the efficacy of our proposed trajectory designs. Specifically, the proposed designs are shown to significantly outperform the simple straight flight trajectory in terms of achievable quality-of-connectivity, and the performance gain becomes more pronounced as the GBS density increases. Furthermore, the proposed designs are shown to achieve close performance to the optimal trajectory obtained via exhaustive search, yet with substantially reduced complexity.
\end{itemize}

\subsection{Organization}
The rest of this paper is organized as follows. Section \ref{sec_system} presents the system model and problem formulation. For the formulated problem, Section \ref{sec_feasibility} proposes a graph connectivity based approach for checking its feasibility. Section \ref{sec_solution} reveals the structural properties of the optimal solution, and proposes two efficient methods based on graph theory for finding high-quality approximate solutions. Numerical results are provided in Section \ref{sec_numerical}. Finally, Section \ref{sec_conclusion} concludes the paper and points out several promising directions for future work.

\textit{Notations}: Scalars and vectors are denoted by lower-case letters and boldface lower-case letters, respectively. ${\mv{z}}^T$ and $\|{\mv{z}}\|$ denote the transpose and the Euclidean norm of a vector $\mv{z}$, respectively. $|\mathcal{X}|$ denotes the cardinality of a set $\mathcal{X}$. $\mathcal{X}\cup\mathcal{Y}$ denotes the union of two sets $\mathcal{X}$ and $\mathcal{Y}$. $\mathcal{O}(\cdot)$ denotes the standard big-O notation. $\mathbb{R}^{m\times n}$ denotes the space of $m\times n$ real matrices. $\mathbb{C}$ denotes the space of complex numbers. $\angle(\cdot)$ denotes the phase of a complex number. The symbol $j$ represents the imaginary unit of complex numbers, i.e., $j=\sqrt{-1}$. The distribution of a circularly symmetric complex Gaussian (CSCG) random variable with mean $\mu$ and variance $\sigma^2$ is denoted by $\mathcal{CN}(\mu,\sigma^2)$; and $\sim$ stands for ``distributed as''. $\mathbb{E}[\cdot]$ denotes the statistical expectation. For a time-dependent function ${\mv{x}}(t)$, $\dot{\mv{x}}(t)$ denotes its first-order derivative {\hbox{with respect to time $t$.}}

\section{System Model and Problem Formulation}\label{sec_system}
As shown in Fig. \ref{cellularUAV}, we consider a cellular-enabled UAV communication system with $M$ GBSs and a UAV flying from a pre-determined initial location $U_0$ to final location $U_F$.\footnote{Without loss of generality, we consider a total number of $M$ GBSs that are available to communicate with the UAV along its trajectory, while in general a larger $M$ is needed as the distance between $U_0$ and $U_F$ increases.} For simplicity, we assume that the UAV flies at a constant altitude of $H$ in meter (m), where $H$ corresponds to the practical constraint on the minimum UAV flying altitude imposed by government regulations for safety consideration. With a 3D Cartesian coordinate system, we denote $(a_m,b_m,H_{\mathrm{G}})$ as the coordinate of the $m$th GBS, where for simplicity we assume that all GBSs have the same altitude $H_{\mathrm{G}}$; while our results can be easily extended to the case with different GBS altitudes. Denote $(x_0,y_0,H)$ and $(x_F,y_F,H)$ as the coordinates of $U_0$ and $U_F$, respectively, and $(x(t),y(t),H)$, $0\leq t\leq T$, as the time-varying coordinate of the UAV, where $T$ denotes the mission completion time, which is determined by the UAV trajectory to be designed. We further define ${\mv{g}}_m=[a_m,b_m]^T$, ${\mv{u}}_0=[x_0,y_0]^T$, ${\mv{u}}_F=[x_F,y_F]^T$ and ${\mv{u}}(t)=[x(t),y(t)]^T$ to represent the above locations projected on the horizontal ground plane, respectively, where ${\mv{u}}(0)={\mv{u}}_0$ and ${\mv{u}}(T)={\mv{u}}_F$.

For the purpose of exposition, we assume that the UAV as well as each GBS is equipped with a single antenna with omnidirectional antenna pattern,\footnote{It is worth noting that when the UAV and/or each GBS are equipped with multiple antennas, our results in this paper are also applicable, by adding a multiplicative beamforming gain to the received SNR given in (\ref{rho}).} and the channel between the UAV and each GBS is dominated by the LoS link, where the Doppler effect due to the UAV mobility is assumed to be compensated perfectly at the receiver. The time-varying distance between the $m$th GBS and the UAV can be expressed as
\begin{equation}\label{distance}
d_{m}(t)=\sqrt{(H-H_{\mathrm{G}})^2+\|{\mv{u}}(t)-{\mv{g}}_{m}\|^2},\quad
m\in \mathcal{M},
\end{equation}
where $\mathcal{M}=\{1,...,M\}$ denotes the set of GBSs. In this paper, we consider the scenario of downlink transmission from GBSs to the UAV, as illustrated in Fig. \ref{cellularUAV}; while our results are also applicable to the uplink transmission from the UAV to GBSs. Let $h_m(t)$ denote the time-varying complex baseband channel coefficient from the $m$th GBS to the UAV. It follows from the LoS channel model that the channel power gain is modeled as
\begin{equation}\label{channel}
|h_{m}(t)|^2=\frac{{\beta_0}}{d_m^2(t)}=\frac{{\beta_0}}{{(H-H_{\mathrm{G}})^2+\|{\mv{u}}(t)-{\mv{g}}_m\|^2}},\quad
m\in \mathcal{M},
\end{equation}
where $\beta_0$ denotes the channel power gain at the reference distance
of $d_0=1$ m \cite{wireless}.

We assume that at each time instant $t$ during the UAV mission, the UAV is associated with one GBS indexed by $I(t)\in\mathcal{M}$ for communication. Let $s\in \mathcal{S}$ denote the information symbol to be sent to the UAV, where $\mathcal{S}$ denotes the symbol constellation with unit average power, i.e., $\mathbb{E}_{s\in\mathcal{S}}[|s|^2]=1$. The received signal at the UAV is thus given by
\begin{align}\label{y}
y(t)=\sqrt{P}h_{I(t)}(t)s+n,\quad 0\leq t\leq T,
\end{align}
where $P$ denotes the transmission power at each GBS; $n\sim \mathcal{CN}(0,\sigma^2)$ denotes the CSCG noise at the UAV receiver with average power $\sigma^2$. It can be observed from (\ref{channel}) and (\ref{y}) that to maximize the received signal power at the UAV, the GBS with the largest channel power gain, namely, the one that is closest to the UAV, should be selected for communication with the UAV, i.e., $I(t)=\arg\underset{m\in \mathcal{M}}{\min}\ \|{\mv{u}}(t)-{\mv{g}}_m\|,\ 0\leq t\leq T$. Consequently, the SNR at the UAV receiver is given by
\begin{equation}
\rho(t)=\frac{\gamma_0}{(H-H_{\mathrm{G}})^2+\underset{m\in
\mathcal{M}}{\min}\
\|{\mv{u}}(t)-{\mv{g}}_m\|^2},\quad 0\leq t\leq T,\label{rho}
\end{equation}
where $\gamma_0=\frac{P\beta_0}{\sigma^2}$ denotes the reference SNR. In this paper, the receiver SNR $\rho(t)$ is taken as the \emph{quality-of-connectivity} of the GBS-UAV communication link. We further consider delay-sensitive communication for the UAV, where a minimum SNR target $\bar{\rho}$
is desired to be achieved at any time instant during the mission, i.e., $\rho(t)$ should satisfy
\begin{align}\label{SNRtarget}
\rho(t)\geq \bar{\rho},\quad 0\leq t\leq T.
\end{align}
Notice from (\ref{rho}) that $\rho(t)$ is determined by the UAV trajectory ${\mv{u}}(t)$, which thus
needs to be properly designed under the SNR constraint given in (\ref{SNRtarget}).

We aim to minimize the UAV mission completion time $T$ by optimizing the UAV trajectory ${\mv{u}}(t)$, subject to the minimum SNR constraint given in (\ref{SNRtarget}), under the assumption that the locations of all GBSs are known \emph{a priori}. Furthermore, denote $V_{\max}$ as the maximum UAV speed. We thus have the additional constraint $\|\dot{\mv{u}}(t)\|\leq V_{\max},\ 0\leq t\leq T$, where $\dot{\mv{u}}(t)$ represents the UAV velocity. By explicitly expressing $\rho(t)$ according to (\ref{rho}), the minimum SNR constraint can be shown to be satisfied if and only if at any time instant during the UAV mission, the horizontal distance between the UAV and its closest GBS, i.e., $\underset{m\in\mathcal{M}}{\min}\ \|{\mv{u}}(t)-{\mv{g}}_m\|$, is no larger than a threshold given by
\begin{align}\label{dtarget}
\bar{d}\overset{\Delta}{=}\sqrt{\tfrac{\gamma_0}{\bar{\rho}}-(H-H_\mathrm{G})^2}.
\end{align}
Note that a smaller $\bar{d}$ corresponds to a larger SNR target $\bar{\rho}$, and hence a more stringent requirement on the quality-of-connectivity. Thus, the optimization problem is formulated as
\begin{align}
\mbox{(P1)}\quad \underset{T,\{{\mv{u}}(t),\ 0\leq t\leq T\}}{\min}\quad
&T\\
\mathrm{s.t.}\quad & {\mv{u}}(0)={\mv{u}}_0\label{P1c_i}\\
& {\mv{u}}(T)={\mv{u}}_F\label{P1c_f}\\
& \underset{m\in \mathcal{M}}{\min}\ \|{\mv{u}}(t)-{\mv{g}}_m\|\leq
\bar{d},\quad 0\leq t\leq
T\label{P1c_SNR}\\
& \|\dot{\mv{u}}(t)\|\leq V_{\max},\quad 0\leq t\leq T.\label{P1c_v}
\end{align}
Note that Problem (P1) is a non-convex optimization problem, since the left-hand side (LHS) of each constraint in (\ref{P1c_SNR}) is the pointwise minimum of a set of convex functions, thus being a non-convex function in general. Moreover, ${\mv{u}}(t)$ is a continuous function of $t$, thus (P1) essentially involves an infinite number of optimization variables. Therefore, the optimal solution to (P1) is in general challenging to obtain.

In the following, we develop a \emph{graph theory based design framework} for solving (P1). Specifically, by exploiting the unique structure of (P1), we establish graph based reformulations of (P1), which can be solved by applying powerful algorithms in graph theory. Under this framework,
we first propose a graph connectivity based approach for checking the feasibility of (P1) in Section \ref{sec_feasibility}. Then, we propose two efficient methods based on graph theory and convex optimization for finding high-quality approximate solutions to (P1) in Section \ref{sec_solution}.

\section{Feasibility Check for Problem (P1): A Graph Connectivity Based Approach}\label{sec_feasibility}
In this section, we study the feasibility of Problem (P1). Note that due to the non-convex constraints in (\ref{P1c_SNR}) and the continuous variable $\{{\mv{u}}(t),\ 0\leq t\leq T\}$, checking the feasibility of (P1) is a difficult task in general. In the following, we propose an efficient \emph{graph connectivity} based approach, by examining all feasible sequences of GBS-UAV associations along the UAV trajectory, which satisfy the constraints in (\ref{P1c_SNR}).

Specifically, notice that with any given UAV trajectory ${\mv{u}}(t)$, the constraints in (\ref{P1c_SNR}) are satisfied if and only if there exists a sequence of GBSs that are successively associated with the UAV over the time horizon $[0,T]$, with the horizontal distance between the UAV and its associated GBS no greater than $\bar{d}$ at any time instant $t\in [0,T]$. Thus, we introduce an auxiliary vector ${\mv{I}}=[I_1,...,I_N]^T$ with $I_i\in \mathcal{M},\ \forall i$ to represent the \emph{GBS-UAV association sequence}, which indicates that the UAV is first associated with GBS $I_1$, and then handed over to GBS $I_2$, etc., with $N-1$ denoting the total number of \emph{GBS handovers}. We then have the following proposition.
\begin{proposition}\label{prop_feas}
Problem (P1) is feasible if and only if there exists a GBS-UAV association sequence ${\mv{I}}=[I_1,...,I_N]^T$ that satisfies the following conditions:
\begin{align}
\|{\mv{u}}_0-{\mv{g}}_{I_1}\|&\leq\bar{d}\label{feas_i}\\
\|{\mv{u}}_F-{\mv{g}}_{I_N}\|&\leq \bar{d}\label{feas_f}\\
\|{\mv{g}}_{I_{i+1}}-{\mv{g}}_{I_i}\|&\leq2\bar{d},\quad
i=1,...,N-1\label{feas_GBS}\\
I_i&\in\mathcal{M},\quad i=1,...,N.\label{feas_I}
\end{align}
\end{proposition}
\begin{IEEEproof}
Please refer to Appendix \ref{proof_prop_feas}.
\end{IEEEproof}

For convenience, we define the (horizontal) \emph{coverage area} of each $m$th GBS under a given quality-of-connectivity requirement specified by $\bar{d}$ as the disk region on the horizontal plane centered at ${\mv{g}}_m$ with radius $\bar{d}$, since the quality-of-connectivity requirement can be always satisfied by associating this GBS with the UAV as long as its horizontal location lies in this region. Then, the results in Proposition \ref{prop_feas} indicate that Problem (P1) is feasible if and only if there exists a GBS-UAV association sequence ${\mv{I}}$, such that the horizontal locations of $U_0$ and $U_F$ lie in the coverage area of the first-associated GBS and the last-associated GBS, respectively; while the intersection between the coverage areas of any two consecutively associated GBSs is non-empty.

\begin{figure}[t]
\centering
  \subfigure[Horizontal locations of $U_0$, $U_F$ and $M=10$ GBSs]{
  \includegraphics[width=12cm]{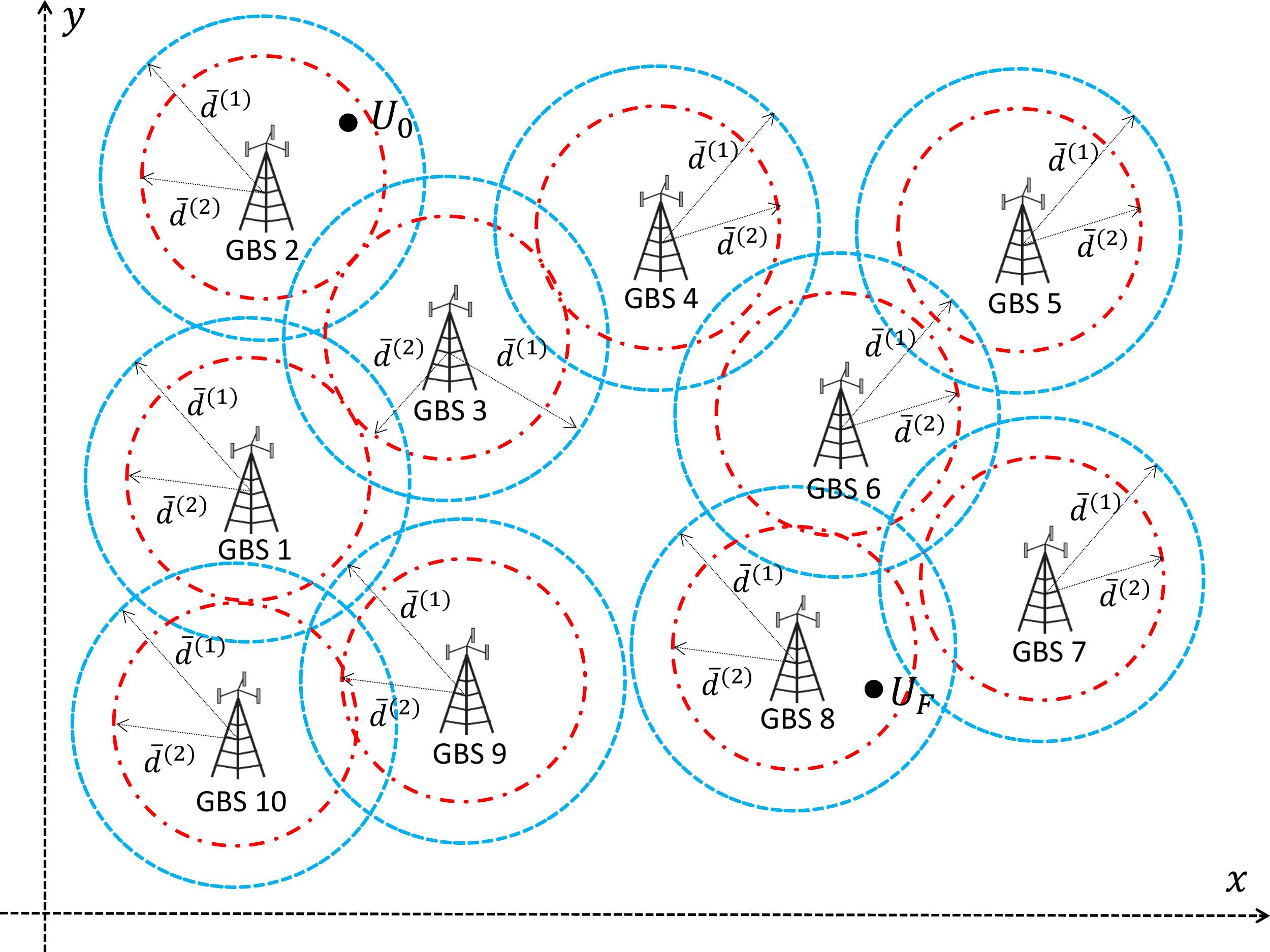}}
  \subfigure[Graph ${G}$ with $\bar{d}=\bar{d}^{(1)}$: Feasible case]{
  \includegraphics[width=6cm]{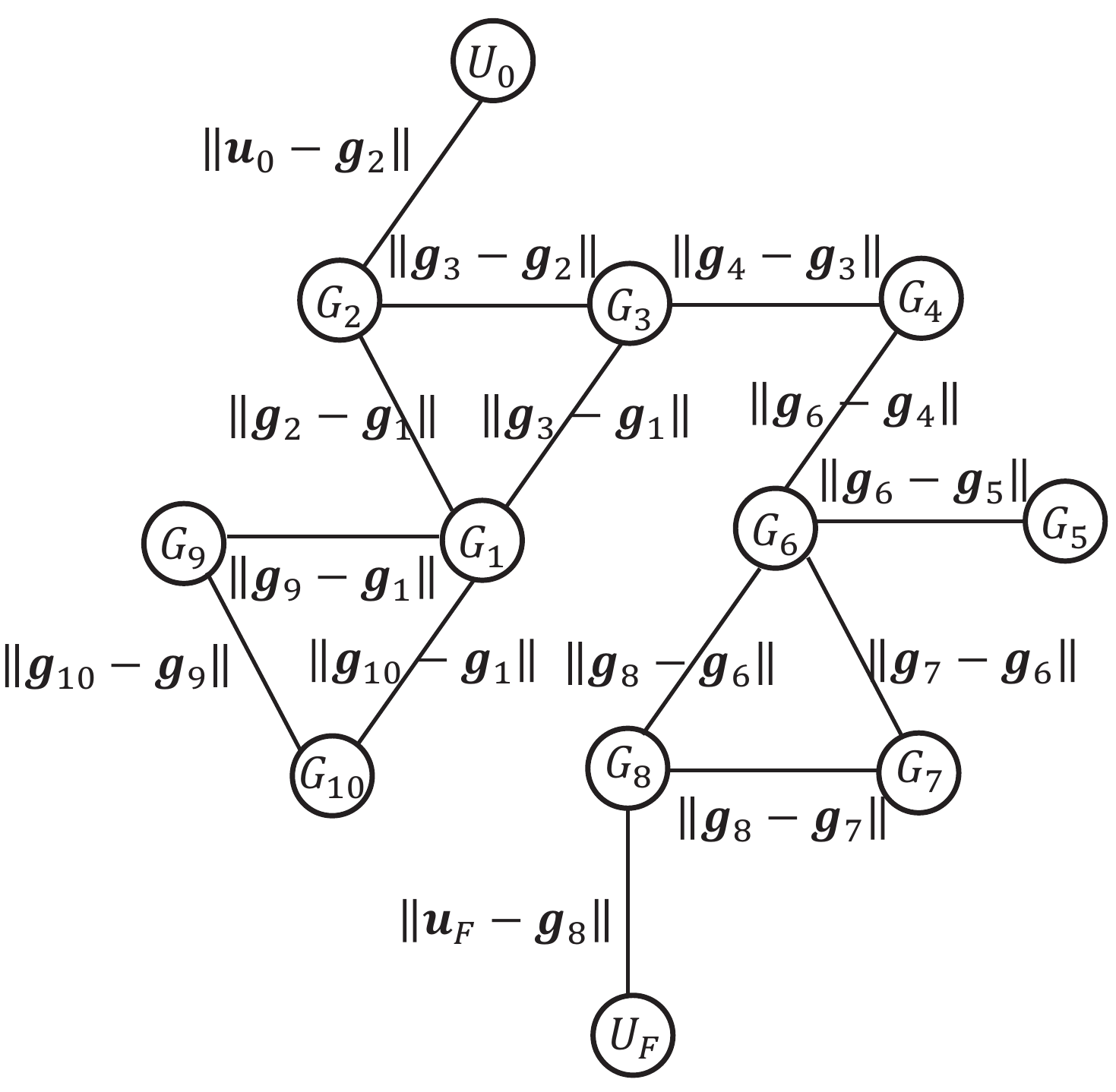}}
  \subfigure[Graph ${G}$ with $\bar{d}=\bar{d}^{(2)}$: Infeasible case]{
  \includegraphics[width=6cm]{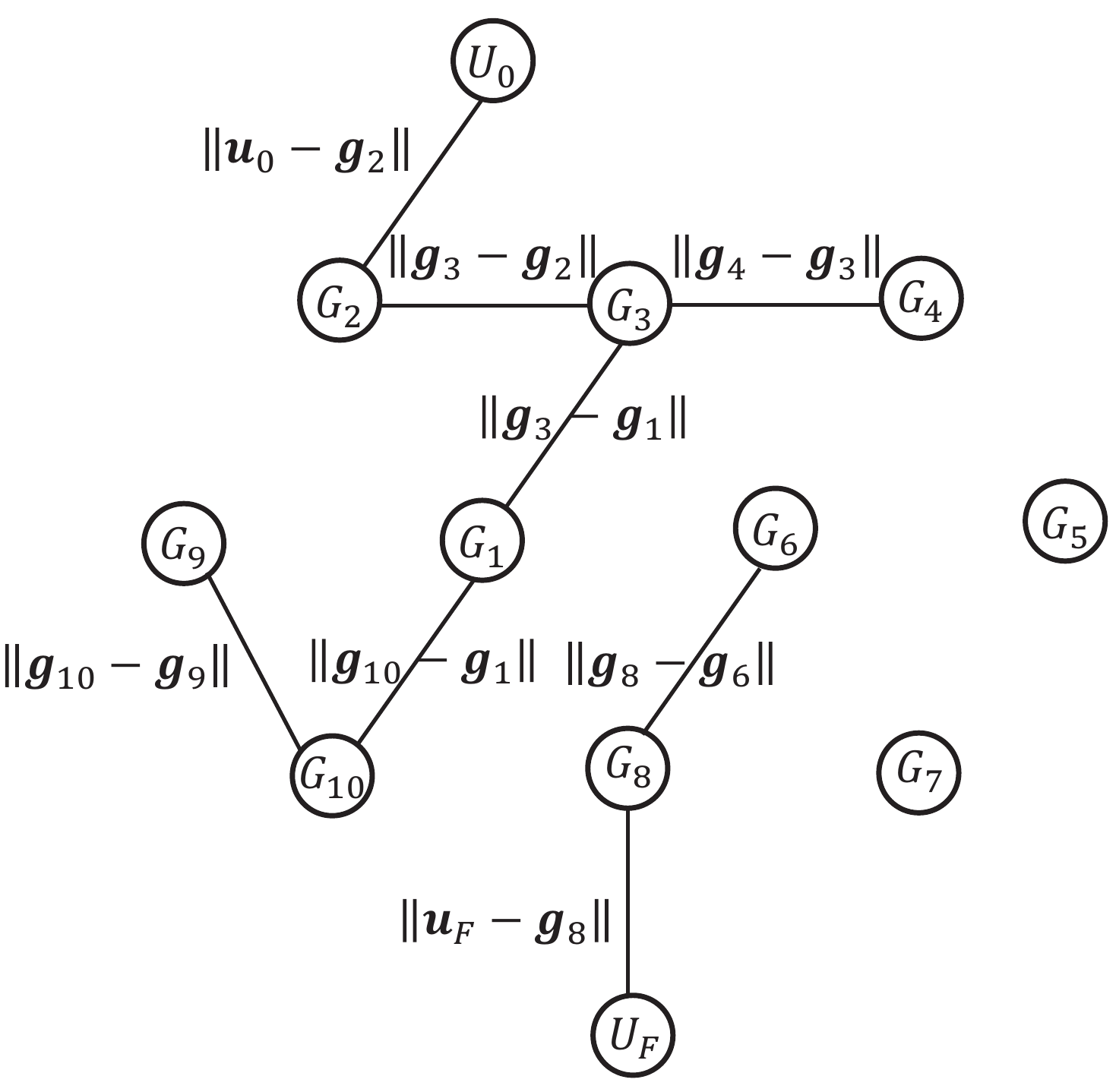}}
  \caption{Illustration of feasibility check for Problem (P1) based on graph connectivity.}
  \label{fig_feas}
\end{figure}

Based on the results in Proposition \ref{prop_feas}, the feasibility of Problem (P1) can be checked
via the following procedure by leveraging \emph{graph connectivity}. First, we construct an undirected weighted \emph{graph} denoted by ${G}=(V,E)$, where the vertex set $V$ is given by
\begin{equation}\label{vertex}
V=\{U_0,G_1,...,G_M,U_F\},
\end{equation}
where $U_0$ and $U_F$ represent the UAV initial and final locations, respectively, and $G_m$ represents the $m$th GBS; the edge set $E$ is given by
\begin{align}\label{edge}
E=&\{(U_0,G_m):\|{\mv{u}}_0-\!{\mv{g}}_m\|\leq \bar{d},\ m\in \mathcal{M}\}\nonumber\\
\cup&\{(G_m,G_n):\|{\mv{g}}_m-{\mv{g}}_n\|\leq 2\bar{d},\ m,n\in \mathcal{M},m\neq n\}\nonumber\\
\cup&\{(U_F,G_m):\|{\mv{u}}_F-{\mv{g}}_m\|\leq \bar{d},\ m\in \mathcal{M}\}.
\end{align}
The weight of each edge is given by
\begin{align}\label{weight}
&W(U_0,G_m)=\|{\mv{u}}_0-{\mv{g}}_m\|,\nonumber\\
&W(U_F,G_m)=\|{\mv{u}}_F-{\mv{g}}_m\|,\nonumber\\
&W(G_m,G_n)=\|{\mv{g}}_m-{\mv{g}}_n\|,\quad m,n\in \mathcal{M},m\neq n.
\end{align}
Note that an edge $(U_0,G_m)$ or $(U_F,G_m)$ exists if and only if the horizontal location of $U_0$ or $U_F$ lies in the coverage area of GBS $m$, i.e., the horizontal distance between $U_0$ or $U_F$ and GBS $m$ is no larger than $\bar{d}$, whose weight is given by the corresponding distance. In addition, an edge $(G_m,G_n)$ exists if and only if the coverage areas of GBSs $m$ and $n$ overlap, i.e., the distance between GBSs $m$ and $n$ is no larger than $2\bar{d}$, whose weight is given by their distance. For illustration, we consider an example system with horizontal locations of $U_0$, $U_F$ and $M=10$ GBSs shown in Fig. \ref{fig_feas}(a). We then show in Fig. \ref{fig_feas}(b) and Fig. \ref{fig_feas}(c) the corresponding graph $G$ to two different values of $\bar{d}$ given by $\bar{d}^{(1)}$ and $\bar{d}^{(2)}=\frac{3}{4}\bar{d}^{(1)}$, respectively, {\hbox{which are illustrated in Fig. \ref{fig_feas}(a).}}

It follows from Proposition \ref{prop_feas} and the definition of ${G}$ that Problem (P1) is feasible if and only if $U_0$ and $U_F$ are \emph{connected} in the constructed graph $G$ with given $\bar{d}$, i.e., ${G}$ contains a \emph{path} from $U_0$ to $U_F$ \cite{graph}. The connectivity between $U_0$ and $U_F$ can be readily verified via various existing algorithms in graph theory, e.g., breadth-first search, with complexity of $\mathcal{O}(M)$ \cite{graph}. Note that constructing the graph ${G}$ also requires a complexity of $\mathcal{O}(M^2)$. Thus, the total complexity for checking the feasibility of Problem (P1) is $\mathcal{O}(M^2)$. As an example, it can be observed that $U_0$ and $U_F$ are connected in the graph shown in Fig. \ref{fig_feas}(b) (e.g., a path can be found as $(U_0,G_2,G_3,G_4,G_6,G_8,U_F)$, which corresponds to a feasible GBS-UAV association sequence ${\mv{I}}=[2,3,4,6,8]^T$), thus (P1) is feasible with $\bar{d}=\bar{d}^{(1)}$; in contrast, $U_0$ and $U_F$ are not connected in the graph shown in Fig. \ref{fig_feas}(c), namely, (P1) is infeasible with $\bar{d}=\bar{d}^{(2)}=\frac{3}{4}\bar{d}^{(1)}$, due to the more stringent connectivity constraints in (\ref{P1c_SNR}) with smaller $\bar{d}$.

\section{Proposed Solution to Problem (P1): Optimal Structure and Two Graph Based Methods}\label{sec_solution}
In this section, we propose high-quality approximate solutions to Problem (P1) assuming that it has been verified to be feasible. First, under the GBS-UAV sequential association model established in Section \ref{sec_feasibility}, we show that the optimal UAV trajectory is determined by the optimal GBS-UAV association sequence as well as the UAV's optimal locations where it is handed over from one GBS to another in the optimal association sequence. Hence, Problem (P1) can be equivalently reformulated into a joint optimization problem of the GBS-UAV association sequence and UAV handover locations, which has a finite number of variables as compared to (P1) that has infinite number of variables. Next, we investigate the joint GBS-UAV association and handover location optimization problem, which is however still non-convex and difficult to be optimally solved. To tackle this difficulty, we first reveal some useful structural properties for its optimal solution. Based on them, we then propose two efficient methods to find approximate solutions for this problem, both of which require only \emph{polynomial complexity} in terms of $M$ and can achieve the optimal value within a \emph{bounded gap}, by applying graph theory and convex optimization techniques. Finally, we summarize our developed graph theory based algorithms for Problem (P1).
\subsection{Problem Reformulation Based on GBS-UAV Association Sequence}\label{sec_reformulation}
To start with, we reformulate Problem (P1) into a more tractable form by re-expressing the constraints in (\ref{P1c_SNR}) based on a given GBS-UAV association sequence. Specifically, recall from Section \ref{sec_feasibility} that the indices of the GBSs that are sequentially associated with the UAV can be represented by an auxiliary vector ${\mv{I}}=[I_1,...,I_N]^T$, where $I_i\in \mathcal{M}$ denotes the GBS associated with the UAV between the $(i-1)$th and the $i$th handovers. Besides $\mv{I}$, we further introduce a set of auxiliary variables $\{T_i\}_{i=1}^{N}$, where $T_i$ denotes the time duration between the $(i-1)$th and $i$th handovers for $i=2,...,N-1$, $T_1$ denotes the time duration from the mission start to the first handover, and $T_{N}$ denotes the time duration from the $(N-1)$th handover to the mission completion. By leveraging the auxiliary variables ${\mv{I}}$ and $\{T_i\}_{i=1}^{N}$, we have the following proposition.
\begin{proposition}\label{prop_P1eq}
Problem (P1) is equivalent to the following problem:
\begin{align}
{\mbox{(P2)}}\underset{T,\{{\mv{u}}(t),\ 0\leq t\leq T\},{\mv{I}},\{T_i\}_{i=1}^N}{\min} & T\\
\mathrm{s.t.}\quad & (\ref{P1c_i}),(\ref{P1c_f}),(\ref{P1c_v}),(\ref{feas_i}),(\ref{feas_f}),(\ref{feas_GBS}),(\ref{feas_I})\\
& \|{\mv{u}}(t)-{\mv{g}}_{I_i}\|\leq \bar{d},\quad  \forall t\in\bigg[\sum_{j=1}^{i-1}T_j,\sum_{j=1}^i T_j\bigg], \quad i=1,...,N\label{P2c_SNR}\\
& \sum_{i=1}^N T_i=T.\label{P2c_t}
\end{align}
\end{proposition}
\begin{IEEEproof}
Please refer to Appendix \ref{proof_prop_P1eq}.
\end{IEEEproof}

Given the equivalence between Problem (P1) and Problem (P2), we derive a useful structural property of the optimal UAV trajectory in the next subsection.
\subsection{Structural Property of Optimal UAV Trajectory}\label{sec_property}
For convenience, we define the horizontal location of the UAV where it is handed over from GBS $I_i$
to GBS $I_{i+1}$, as
\begin{equation}
{{\mv{u}}^i}\overset{\Delta}{=}{\mv{u}}\bigg(\sum_{j=1}^i T_j\bigg),\quad i=1,...,N-1,
\end{equation}
which is referred to as the $i$th \emph{handover location} of the UAV. Note that under the constraints in (\ref{P1c_SNR}), each $i$th handover point has a horizontal distance no larger than $\bar{d}$ with both GBSs $I_i$ and $I_{i+1}$, i.e., the feasible region of ${\mv{u}}^i$ is given by
\begin{align}\label{Ui}
&\mathcal{U}_i=\{{\mv{u}}^i\in \mathbb{R}^{2\times 1}:\|{\mv{u}}^i-{\mv{g}}_{I_i}\|\leq \bar{d},\ \|{\mv{u}}^{i}-{\mv{g}}_{I_{i+1}}\|\leq \bar{d}\},\quad i=1,...,N-1.
\end{align}
Notice that $\mathcal{U}_i$ is the intersection between the coverage areas of GBSs $I_i$ and $I_{i+1}$. For completeness, we further define ${{\mv{u}}^0}\overset{\Delta}{=}{{\mv{u}}(0)}={\mv{u}}_0$ and ${\mv{u}}^N\overset{\Delta}{=}{\mv{u}}(T)={\mv{u}}_F$ as the $0$th and the $N$th handover locations, respectively.

Next, based on the UAV handover locations $\{{\mv{u}}^i\}_{i=0}^N$ defined above, we reveal a useful structural property of the optimal UAV trajectory for Problem (P2), as shown in the following proposition.
\begin{proposition}[Trajectory with Connected Line Segments and Maximum Speed]\label{prop_P2}
The optimal solution to Problem (P2) satisfies the following conditions:
\begin{align}
T_i&=\frac{\left\|{{\mv{u}}^i}-{{\mv{u}}^{i-1}}\right\|}{V_{\max}},\quad i=1,...,N\label{Ti}\\
{\mv{u}}(t)&={{\mv{u}}^{i-1}}+\bigg(t-\sum_{j=1}^{i-1} T_j\bigg)V_{\max}\frac{{\mv{u}}^{i}-{\mv{u}}^{i-1}}{\|{\mv{u}}^{i}-{\mv{u}}^{i-1}\|},\quad t\in\bigg[\sum_{j=1}^{i-1}T_j,\sum_{j=1}^i T_j\bigg],\ i=1,...,N\label{ut}\\
T&=\sum_{i=1}^N\frac{\left\|{{\mv{u}}^i}-{{\mv{u}}^{i-1}}\right\|}{V_{\max}}.\label{T}
\end{align}
\end{proposition}
\begin{IEEEproof}
Please refer to Appendix \ref{proof_prop_P2}.
\end{IEEEproof}

\begin{figure}[t]
  \centering
  \includegraphics[width=11cm]{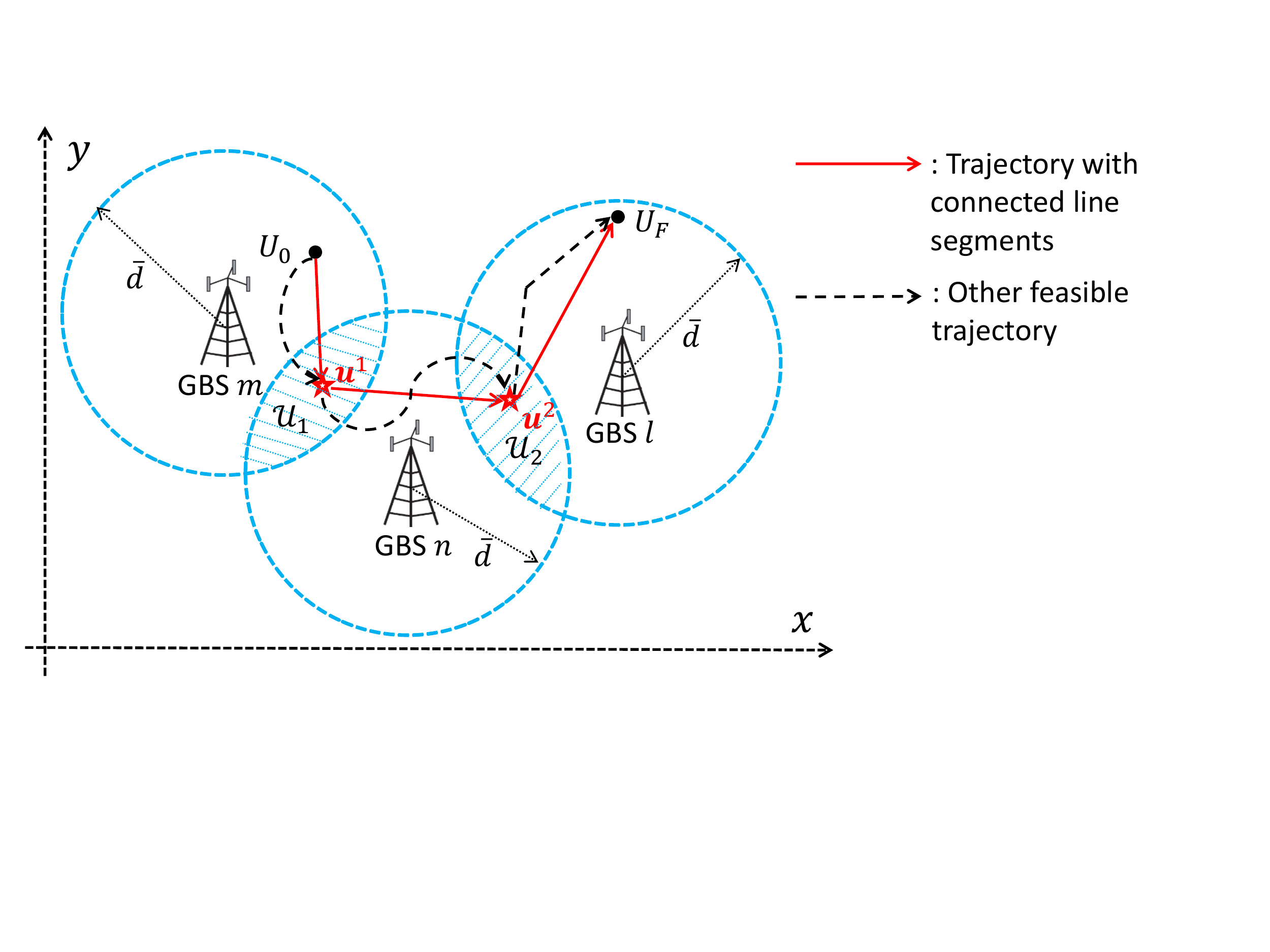}
  \caption{Illustration of the optimal UAV trajectory structure consisting of connected line segments with ${\mv{I}}=[m,n,l]^T$.}\label{Connected_Segments}
\end{figure}

The results in Proposition \ref{prop_P2} indicate that in the optimal solution to Problem (P2) as well as Problem (P1), the UAV should fly from $U_0$ to $U_F$ by following a path consisting of only \emph{connected line segments} and with its \emph{maximum speed}. Moreover, the UAV is associated with the same GBS when it flies along each line segment, and the starting and ending points of each $i$th line segment are the $(i-1)$th and the $i$th handover points with horizontal locations specified by ${\mv{u}}^{i-1}$ and ${\mv{u}}^{i}$, respectively. In Fig. \ref{Connected_Segments}, we illustrate such an optimal UAV trajectory structure consisting of connected line segments only via comparison with other feasible trajectories which in general constitute curved paths of longer distance by assuming ${\mv{I}}=[m,n,l]^T$ and a set of handover locations $\{{\mv{u}}^i\}_{i=1}^{2}$.

By leveraging the above optimal structure, Problem (P2) can be readily shown equivalent to the following problem, which aims to minimize the \emph{total flying distance} of the UAV by jointly optimizing the \emph{GBS-UAV association sequence} ${\mv{I}}$ and the \emph{handover locations} $\{{\mv{u}}^i\}_{i=0}^N$:
\begin{align}
{\mbox{(P3)}}\quad \underset{{\mv{I}},\{{\mv{u}}^i\}_{i=0}^N}{\min}\quad &
\sum_{i=1}^N
\|{\mv{u}}^i-{\mv{u}}^{i-1}\|\\
\mathrm{s.t.}\quad & {\mv{u}}^0={\mv{u}}_0\label{P3c_i}\\
& {\mv{u}}^N={\mv{u}}_F\label{P3c_f}\\
& \|{\mv{u}}^i-{\mv{g}}_{I_i}\|\leq \bar{d},\quad i=1,...,N
\label{P3c_SNR1}\\
& \|{\mv{u}}^{i-1}-{\mv{g}}_{I_{i}}\|\leq \bar{d},\quad i=1,...,N
\label{P3c_SNR2}\\
& (\ref{feas_i}),(\ref{feas_f}),(\ref{feas_GBS}),(\ref{feas_I}).
\end{align}

Notice that by characterizing the continuous UAV trajectory $\{{\mv{u}}(t),\ 0\leq t\leq T\}$ with a discrete set of handover locations $\{{\mv{u}}^i\}_{i=0}^N$, Problem (P3) significantly reduces the number of variables compared to Problem (P2). Since (P2) and (P1) are equivalent as shown in Proposition \ref{prop_P1eq}, (P3) is also equivalent to (P1), whose solution can be readily obtained by substituting the solution to (P3) into (\ref{ut}) and (\ref{T}). Thus, the remaining task is to solve Problem (P3), as addressed next.

\subsection{Joint GBS-UAV Association and Handover Location Optimization}\label{sec_P3}
In this subsection, we aim to solve the joint GBS-UAV association and handover location optimization problem given by Problem (P3). It is worth noting that (P3) is a non-convex combinatorial optimization problem due to the discrete variables in the GBS-UAV association sequence $\mv{I}$. Also, note that the length of $\mv{I}$, i.e., $N$, is also an implicit variable although it is intuitively upper-bounded by $M$ (to be rigorously shown in the following). As a result, exhaustively searching over all possible association sequences of $\mv{I}$ can result in a prohibitive complexity of $\mathcal{O}(M!)$. Moreover, another challenge for solving (P3) lies in the coupling between ${\mv{I}}$ and the handover locations $\{{\mv{u}}^i\}_{i=1}^{N-1}$ through the constraints in (\ref{P3c_SNR1}) and (\ref{P3c_SNR2}). Thus, Problem (P3) is difficult to be optimally solved. In the following, we first derive useful properties of the optimal GBS-UAV association sequence and handover locations for (P3). Then, by leveraging these properties, we propose two effective methods for finding high-quality approximate solutions to (P3) based on graph theory and convex optimization techniques.

\subsubsection{Properties of Optimal GBS-UAV Association and Handover Location Solution}
First, we show one property of the optimal GBS-UAV association sequence $\mv{I}$ in the following proposition, which helps significantly reduce the search space for the optimal $\mv{I}$.
\begin{proposition}[Non-Repeated GBS-UAV Association]\label{lemma_I}
The optimal GBS-UAV association sequence ${\mv{I}}=[I_1,...,I_N]^T$ to Problem (P3) satisfies $I_i\neq I_j,\ \forall i\neq j$ and thus $N\leq M$.
\end{proposition}
\begin{IEEEproof}
Please refer to Appendix \ref{proof_lemma_I}.
\end{IEEEproof}

Proposition \ref{lemma_I} implies that the UAV shall not be associated with the same GBS in two non-consecutive time intervals, and the total number of handovers during the UAV mission is no larger than $M-1$. This is consistent with our intuition that in order to minimize the total flying distance, the UAV shall not return to the neighbourhood of its previously traveled locations. Note that for Problem (P3), the feasible set of ${\mv{I}}$ under the constraints in (\ref{feas_i})--(\ref{feas_I}) constitutes all \emph{walks} from $U_0$ to $U_F$ in the graph $G=(V,E)$ defined in (\ref{vertex})--(\ref{weight}) in Section \ref{sec_feasibility} \cite{graph}. The total number of such walks is infinite since an arbitrary number of repeated edges and/or vertices may exist in each walk. In contrast, based on the results in Proposition \ref{lemma_I}, the optimal set of $\mv{I}$ is composed of all \emph{paths} from $U_0$ to $U_F$ in graph $G$ without repeated edges/vertices, whose cardinality is finite and significantly smaller than that of all feasible walks.

Next, note that for Problem (P3), the feasible region of each $i$th handover location ${\mv{u}}^i$ under the constraints in (\ref{P3c_SNR1}) and (\ref{P3c_SNR2}) can be expressed as $\mathcal{U}_i$ given in (\ref{Ui}). In the following proposition, we show one property of the optimal $\{{\mv{u}}^i\}_{i=1}^{N-1}$, which results in an equivalent optimal region of each ${\mv{u}}^i$ that is generally much smaller than $\mathcal{U}_i$.
\begin{proposition}[Handover on Intersected Coverage Boundary]\label{prop_edge}
Without loss of optimality, the handover locations $\{{\mv{u}}^i\}_{i=1}^{N-1}$ for Problem (P3) can be assumed to satisfy the following conditions:
\begin{align}
{\mv{u}}^i\in&\ \mathcal{E}_i,\quad i=1,...,N-1\label{edge}\\
&\mathrm{where}\ \ \mathcal{E}_i=\{{\mv{u}}^i\in \mathbb{R}^{2\times 1}:\|{\mv{u}}^i-{\mv{g}}_{I_i}\|=\bar{d},\ {\mv{u}}^i\in\mathcal{U}_i\},\quad i=1,...,N-1.\label{Ei}
\end{align}
\end{proposition}
\begin{IEEEproof}
Please refer to Appendix \ref{proof_prop_edge}.
\end{IEEEproof}

\begin{figure}[t]
  \centering
  \includegraphics[width=13cm]{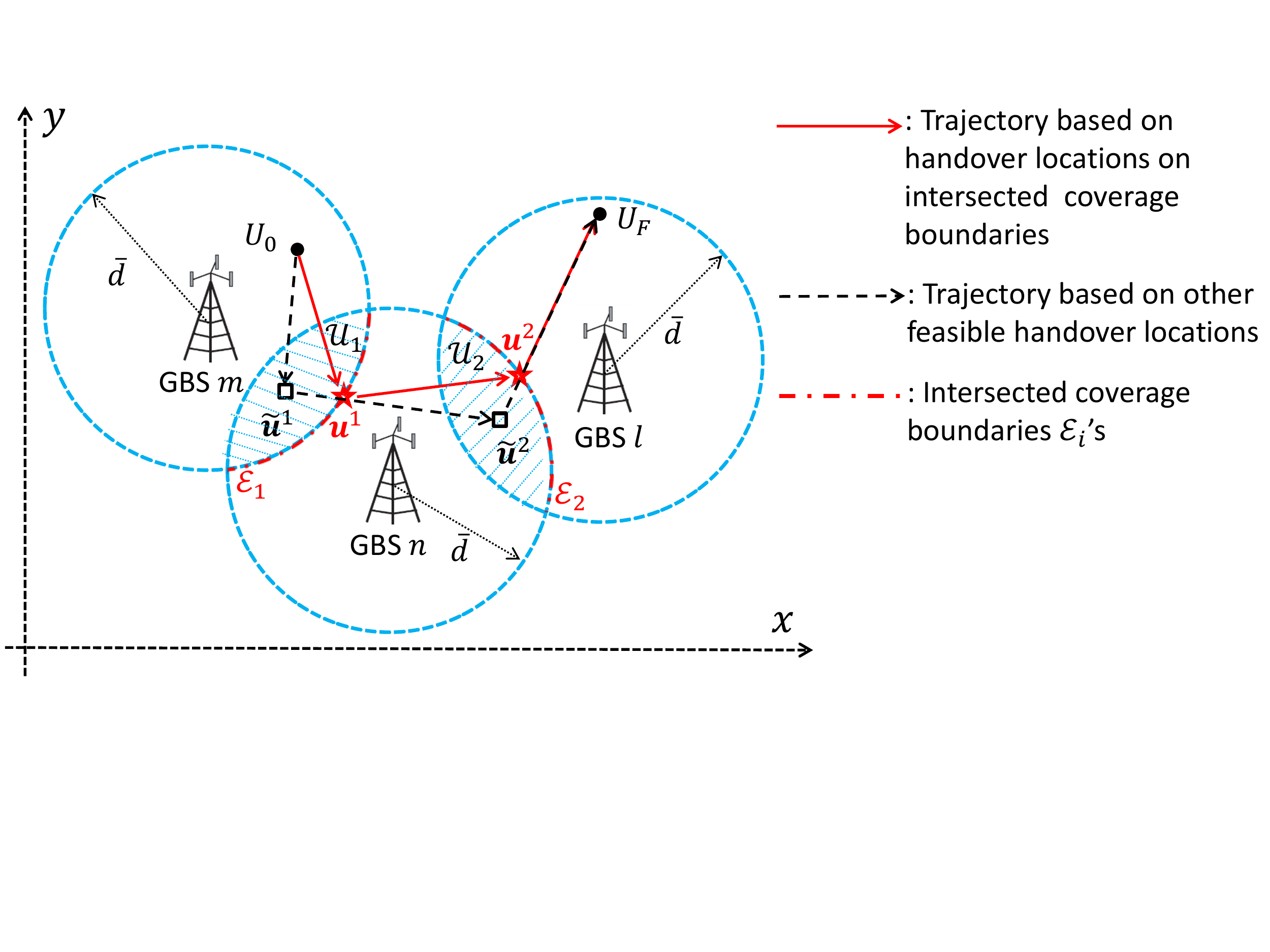}
  \caption{Illustration of the intersected coverage boundaries with ${\mv{I}}=[m,n,l]^T$.}\label{Handover_Edge}
\end{figure}

Recall from Section \ref{sec_feasibility} and Section \ref{sec_property} that the coverage area of GBS $I_i$ is defined as the disk region centered at ${\mv{g}}_{I_i}$ with radius $\bar{d}$, and $\mathcal{U}_i$ is the intersection between the coverage areas of GBSs $I_i$ and $I_{i+1}$. For the purpose of exposition, we further define the \emph{coverage boundary} of a GBS as the circumstance of its disk-shaped coverage area. The results in Proposition \ref{prop_edge} then indicate that the optimal $i$th handover location can be assumed to lie on the intersection between the coverage boundary of GBS $I_i$ and $\mathcal{U}_i$, denoted by $\mathcal{E}_i$, which is further referred to as the ``\emph{intersected coverage boundary}''. In Fig. \ref{Handover_Edge}, we illustrate the intersected coverage boundaries $\mathcal{E}_i$'s by assuming ${\mv{I}}=[m,n,l]^T$.

By leveraging the above properties, we present two efficient methods for finding approximate solutions to Problem (P3) in the following, respectively.

\subsubsection{Method I: Graph Theory Based Association Design and Convex Optimization Based Handover Location Design}
Note that with any given GBS-UAV association sequence $\mv{I}$, Problem (P3) is a convex optimization problem, since the objective function of (P3) is convex over $\{{\mv{u}}^i\}_{i=0}^N$, and the feasible set of each $i$th UAV handover location, $\mathcal{U}_i$, is convex, as can be seen in Fig. \ref{Handover_Edge}. The optimal handover locations with given $\mv{I}$, denoted by $\{{{\mv{u}}^i}^\star({\mv{I}})\}_{i=0}^N$, can thus be efficiently obtained via standard convex optimization techniques or existing software, e.g., CVX \cite{cvx}, with polynomial complexity over $N$ (or $M$) \cite{convex}. Therefore, the optimal solution to Problem (P3) can be obtained by finding $\{{{\mv{u}}^i}^\star({\mv{I}})\}_{i=0}^N$ for all feasible solutions of $\mv{I}$, and selecting the one that yields the minimum objective value (total flying distance). As a result, Problem (P3) reduces to the following GBS-UAV association optimization problem:
\begin{align}
{\mbox{(P3-I)}}\quad \underset{{\mv{I}}}{\min}\quad & \sum_{i=1}^N \|{{\mv{u}}^i}^\star({\mv{I}})-{{\mv{u}}^{i-1}}^\star({\mv{I}})\|\\
\mathrm{s.t.}\quad & I_i\neq I_j, \quad  \forall i\neq j,\ i,j=1,...,N\label{P5c_different}\\
& (\ref{feas_i}),(\ref{feas_f}),(\ref{feas_GBS}),(\ref{feas_I}),
\end{align}
where the constraints in (\ref{P5c_different}) are due to the \emph{non-repeated GBS-UAV association} property given in Proposition \ref{lemma_I}.

It is worth noting that optimally solving Problem (P3) by finding the optimal solution to Problem (P3-I) via \emph{exhaustive search} involves finding all possible paths from $U_0$ to $U_F$ in the graph $G=(V,E)$ defined in (\ref{vertex})--(\ref{weight}) in Section \ref{sec_feasibility}, which requires a prohibitive complexity, e.g., $\mathcal{O}(M!)$ via the depth-first search \cite{graph}, and thus is unaffordable even for moderate $M$. Furthermore, it is difficult to apply conventional optimization techniques for optimally solving (P3-I) because of the following two reasons: i) the objective function of (P3-I) cannot be explicitly expressed, due to the difficulty in obtaining closed-form expressions of $\{{{\mv{u}}^i}^\star({\mv{I}})\}_{i=1}^{N-1}$; ii) (P3-I) is a non-convex combinatorial optimization problem due to the discrete variables $I_i$'s.

Hence, we propose a two-step method for finding an approximate solution to Problem (P3) with the help of Problem (P3-I). In the first step, we derive an upper bound for the objective function of (P3-I) in a tractable form with respect to ${\mv{I}}$, with the aim of minimizing this upper bound subject to the constraints in (\ref{P5c_different}) and (\ref{feas_i})--(\ref{feas_I}). Based on graph $G$ constructed in Section \ref{sec_feasibility} for the feasibility check of Problem (P1), we model this new problem as a classic \emph{shortest path problem} in $G$, whose optimal solution, denoted as $\tilde{\mv{I}}$, can be efficiently obtained with polynomial complexity of $\mathcal{O}(M^2)$ and serves as an approximate solution to (P3-I). Then, in the second step, we obtain the optimal handover locations to Problem (P3) with given $\tilde{\mv{I}}$ denoted by $\{{{\mv{u}}^i}^\star(\tilde{\mv{I}})\}_{i=0}^N$ as previously discussed.

\begin{figure}[t]
  \centering
  \includegraphics[width=12cm]{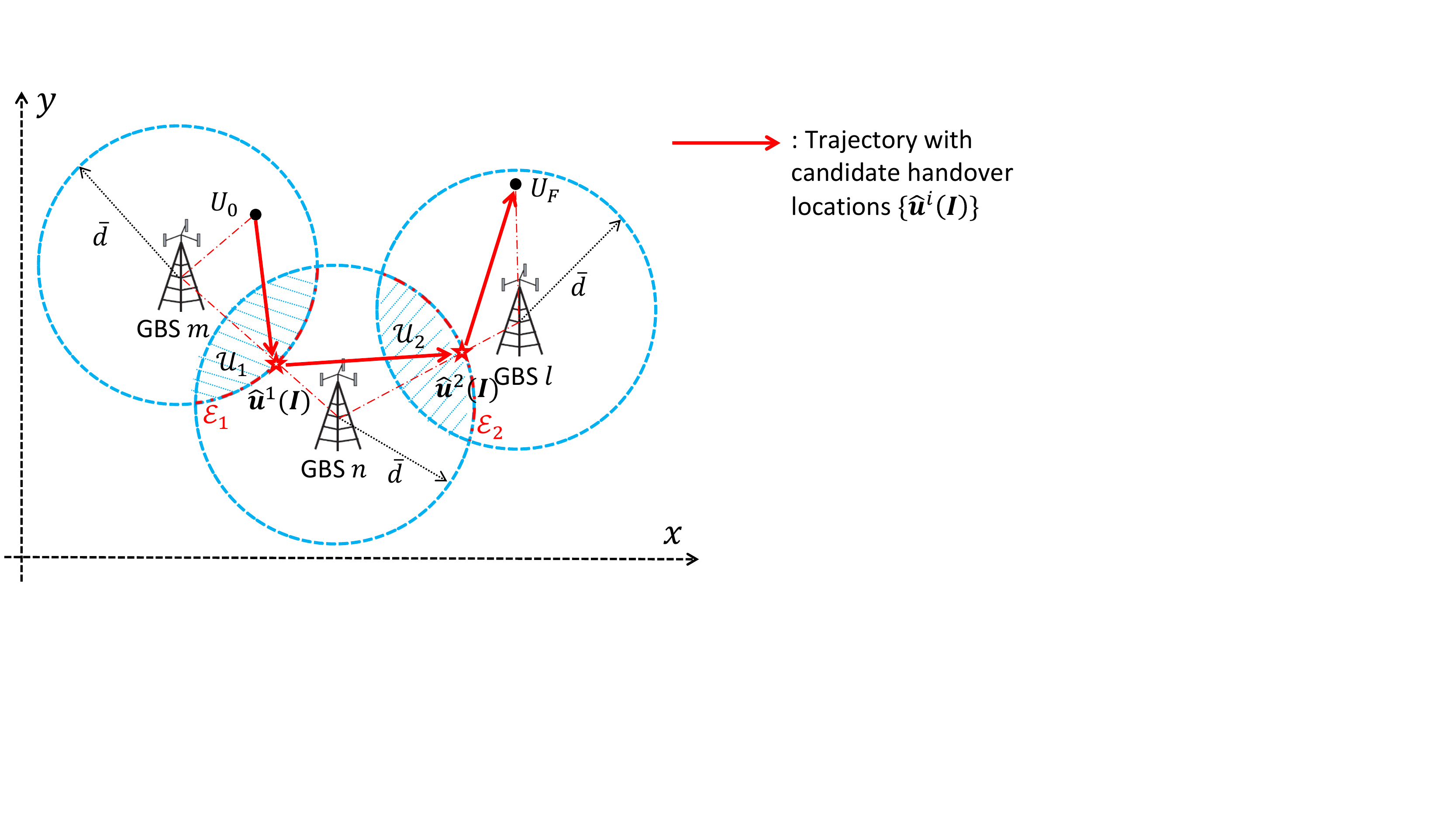}
  \caption{Illustration of a candidate solution of handover locations $\{\hat{\mv{u}}^i({\mv{I}})\}_{i=1}^{N-1}$ with ${\mv{I}}=[m,n,l]^T$.}\label{handover}
\end{figure}
Specifically, note that the objective function of (P3-I), denoted as $s^\star({\mv{I}})\overset{\Delta}{=}\sum_{i=1}^N \|{{\mv{u}}^{i}}^{\star}({\mv{I}})-{{\mv{u}}^{i-1}}^{\star}({\mv{I}})\|$, is upper-bounded by the UAV flying distance with any feasible $\{{\mv{u}}^i\}_{i=0}^N$. Recall from Proposition \ref{prop_edge} that the equivalent optimal region of each ${\mv{u}}^i$, $\mathcal{E}_i$, is the intersected coverage boundary of GBS $I_i$ with the coverage area of GBS $I_{i+1}$. Based on $\mathcal{E}_i$'s, we consider a candidate solution of $\{{\mv{u}}^i\}_{i=0}^N$, denoted as $\{\hat{\mv{u}}^i({\mv{I}})\}_{i=0}^N$, which is given by $\hat{\mv{u}}^0({\mv{I}})={\mv{u}}_0$, $\hat{\mv{u}}^N({\mv{I}})={\mv{u}}_F$, and
\begin{equation}\label{feas_handover}
\hat{\mv{u}}^i({\mv{I}})={\mv{g}}_{I_i}+\bar{d}\frac{{\mv{g}}_{I_{i+1}}-{\mv{g}}_{I_{i}}}{\|{\mv{g}}_{I_{i+1}}-{\mv{g}}_{I_{i}}\|},\quad
i=1,...,N-1.
\end{equation}
Note that each handover location $\hat{\mv{u}}^i({\mv{I}})$ in (\ref{feas_handover}) is the intersecting point of the line segment between the horizontal locations of its consecutively associated two GBSs and the coverage boundary of its formerly associated GBS, namely, the central point of $\mathcal{E}_i$, as illustrated in Fig. \ref{handover}. It then follows that
\begin{align}\label{upp}
s^\star({\mv{I}})=&\sum_{i=1}^N
\|{{\mv{u}}^{i}}^{\star}({\mv{I}})-{{\mv{u}}^{i-1}}^{\star}({\mv{I}})\|\leq \sum_{i=1}^N \|{\hat{\mv{u}}^{i}}({\mv{I}})-{\hat{\mv{u}}^{i-1}}({\mv{I}})\|\nonumber\\
\overset{(a_1)}{\leq}& \|{\mv{u}}_0-{\mv{g}}_{I_1}\|+\sum_{i=1}^{N-1} \|{\mv{g}}_{I_{i+1}}-{\mv{g}}_{I_i}\|+\|{\mv{u}}_F-{\mv{g}}_{I_N}\|\overset{\Delta}{=}\hat{s}({\mv{I}}),
\end{align}
where $(a_1)$ holds due to the triangle inequality, as illustrated in Fig. \ref{handover}. Note that the above upper bound of $s^\star({\mv{I}})$, denoted as $\hat{s}({\mv{I}})$, represents the flying distance of the UAV when it sequentially traverses over the associated GBSs during its mission, which is equivalent to the sum edge weight of the path from $U_0$ to $U_F$ specified by $\mv{I}$ in graph $G=(V,E)$ defined in (\ref{vertex})--(\ref{weight}) in Section \ref{sec_feasibility}. Therefore, finding the optimal ${\mv{I}}$ that minimizes $\hat{s}({\mv{I}})$ subject to the constraints in (\ref{P5c_different}) and (\ref{feas_i})--(\ref{feas_I}) is equivalent to finding the \emph{shortest path} from $U_0$ to $U_F$ in graph ${G}$, which can be efficiently obtained via various existing algorithms with low complexity, e.g., the Dijkstra algorithm with complexity $\mathcal{O}(M^2)$ \cite{graph}.

The above algorithm is summarized as Algorithm \ref{algo_P3_M1} for finding an approximate solution to Problem (P3) with Method I. In the following, we evaluate the complexity and performance of Method I analytically. Note that the complexity for constructing graph $G$ and that for implementing the Dijkstra algorithm have been shown to be both of $\mathcal{O}(M^2)$. In addition, Problem (P3) with given $\mv{I}$ can be cast as a second-order cone program (SOCP) \cite{SOCP}, which can be solved via the interior point method with complexity $\mathcal{O}(N^{3.5})$ \cite{SOCP}. Hence, the overall complexity for Algorithm \ref{algo_P3_M1} can be shown to be no larger than $\mathcal{O}(M^{3.5})$ since $N\leq M$ holds according to Proposition \ref{lemma_I}. On the other hand, let $s^\star$ denote the optimal value of (P3), and $\tilde{s}^\mathrm{I}$ denote the objective value of (P3) with the proposed solution $(\tilde{\mv{I}},\{\tilde{{\mv{u}}}^i\}_{i=0}^{\tilde{N}})$. Then, the performance gap between the proposed solution and the optimal solution is bounded as follows:
\begin{align}\label{bound1}
0\leq\tilde{s}^\mathrm{I}-s^\star\overset{(a_2)}{\leq} \hat{s}(\tilde{\mv{I}})-s^\star\overset{(a_3)}{\leq} 2M\bar{d}-s^\star\overset{(a_4)}{\leq} 2M\bar{d},
\end{align}
where $(a_2)$ holds since $\tilde{s}^\mathrm{I}\leq \hat{s}(\tilde{\mv{I}})$ due to the optimization of handover locations; $(a_3)$ follows from the definition of $\hat{s}({\mv{I}})$ in (\ref{upp}) by using $\|{\mv{u}}_0-{\mv{g}}_{\tilde{I}_1}\|\leq \bar{d}$, $\|{\mv{u}}_F-{\mv{g}}_{\tilde{I}_{\tilde{N}}}\|\leq \bar{d}$, $\|{\mv{g}}_{\tilde{I}_{i+1}}-{\mv{g}}_{\tilde{I}_{i}}\|\leq 2\bar{d}$, and $\tilde{N}\leq M$; and $(a_4)$ holds since $s^\star\geq 0$.
\begin{algorithm}[t]\label{algo_P3_M1}
\caption{Proposed Algorithm for Problem (P3) with Method I}
\SetKwData{Index}{Index}
\KwIn{$\bar{d}$, ${\mv{u}}_0$, ${\mv{u}}_F$, $\{{\mv{g}}_m\}_{m=1}^{M}$, $H$, $H_G$, $V_{\max}$}
\KwOut{$\tilde{\mv{I}}$, $\{\tilde{\mv{u}}^i\}_{i=0}^{\tilde{N}}$}
Construct a graph ${G}=(V,E)$ based on (\ref{vertex})--(\ref{weight}).\\
Find the shortest path from $U_0$ to $U_F$ in ${G}$ via Dijkstra algorithm, and denote the path as $(U_0,G_{\tilde{I}_1},...,G_{\tilde{I}_{\tilde{N}}},U_F)$. Obtain $\tilde{\mv{I}}=[\tilde{I}_1,...,\tilde{I}_{\tilde{N}}]^T$.\\
Obtain $\{{{\mv{u}}^i}^\star(\tilde{\mv{I}})\}_{i=0}^{\tilde{N}}$ by solving Problem (P3) with given
$\tilde{\mv{I}}$ via convex optimization.\\
Obtain $\tilde{\mv{u}}^i={{\mv{u}}^i}^\star(\tilde{\mv{I}}),\ i=0,...,\tilde{N}$.
\end{algorithm}

\subsubsection{Method II: Graph Theory Based Joint Association and Handover Location Design}
\begin{figure}[t]
\centering
  \subfigure[Horizontal locations of $U_0$, $U_F$ and $M=5$ GBSs]{
  \includegraphics[width=8cm]{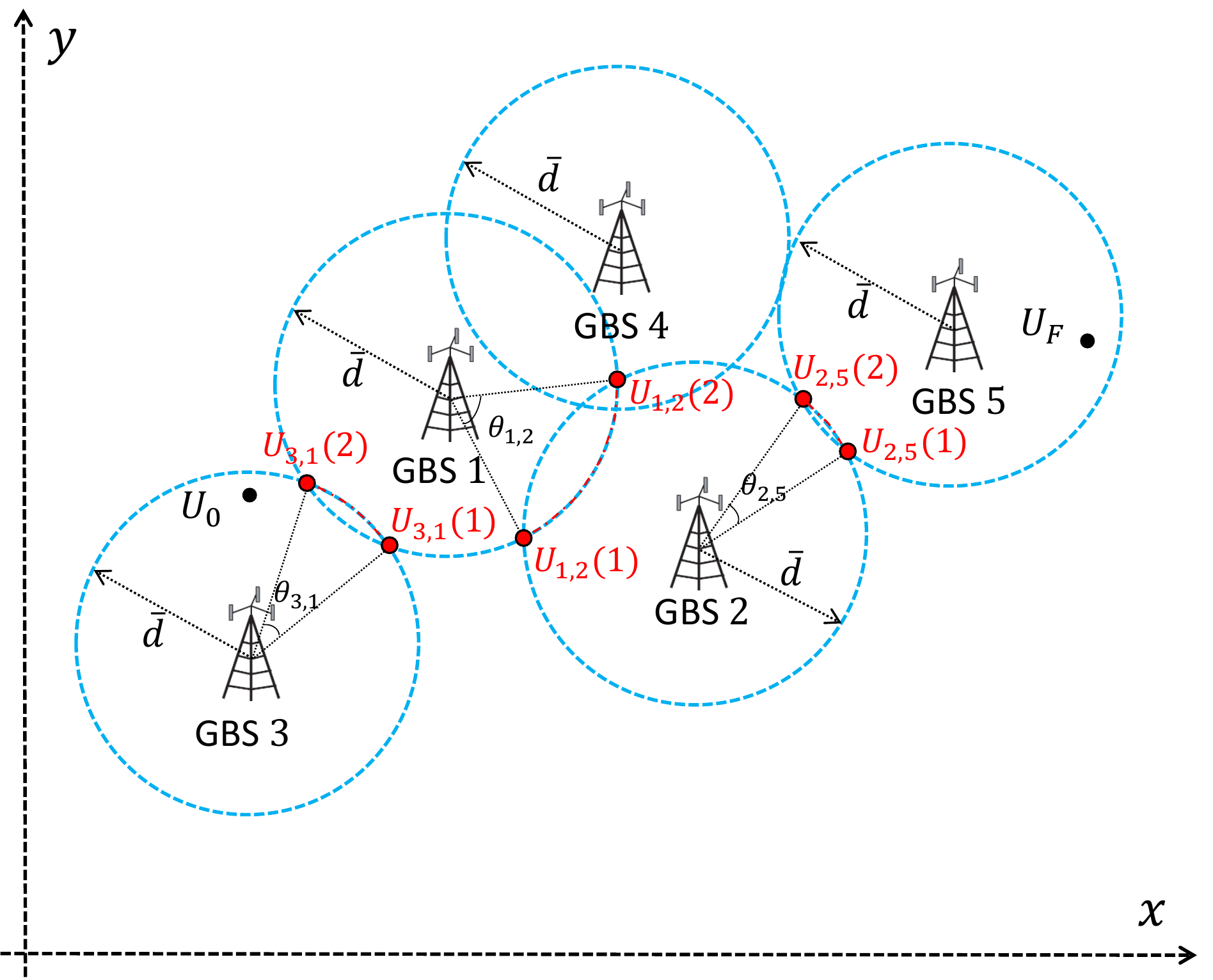}}
  \subfigure[Graph ${G}_Q$ with $Q=2$]{
  \includegraphics[width=8cm]{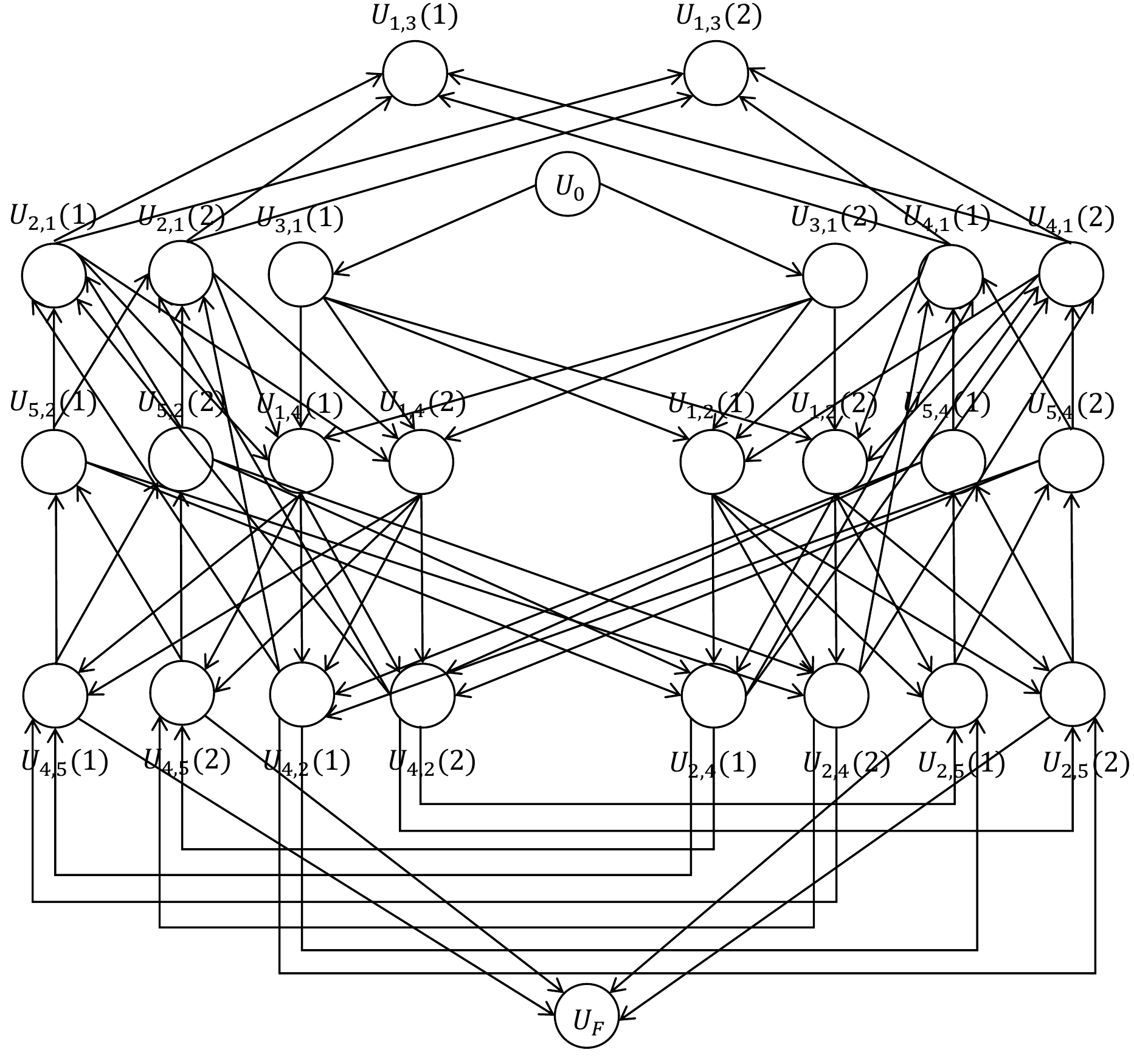}}
  \caption{Illustration of graph construction in Method II for Problem (P3).}
  \label{directed_graph}
\end{figure}

Note that in Method I, the GBS-UAV association and handover location designs are obtained separately in two steps via graph theory and convex optimization techniques, respectively. However, the upper bound in (\ref{upp}) for approximating the objective function of Problem (P3-I) may not be sufficiently tight as can be observed from Fig. \ref{handover}, which may lead to a suboptimal GBS-UAV association design. For example, with the horizontal locations of $U_0$, $U_F$ and $M=5$ GBSs shown in Fig. \ref{directed_graph}(a), the optimal GBS-UAV association sequence by exhaustively searching over graph $G$ is ${\mv{I}}^\star=[3,1,2,5]^T$, while that designed via Method I is $\tilde{\mv{I}}=[3,1,4,5]^T$. This suboptimality occurs because the relationship between a given GBS-UAV association sequence ${\mv{I}}$ and the corresponding optimal handover locations $\{{{\mv{u}}^i}^\star({\mv{I}})\}_{i=0}^N$ is not fully {\hbox{exploited in Method I.}}

In the following, we propose an alternative method for finding an approximate solution to Problem (P3) by jointly designing the GBS-UAV association sequence and handover locations via a different graph construction method, which is able to achieve an arbitrarily small performance gap with the optimal solution at the cost of moderately increased complexity.

First, recall from Proposition \ref{prop_edge} that the optimal $i$th handover location ${\mv{u}}^i$ can be assumed to lie on the intersected coverage boundary of GBS $I_i$ with the coverage area of GBS $I_{i+1}$, as given in (\ref{edge}), (\ref{Ei}) and illustrated in Fig. \ref{Handover_Edge}. Note that compared to its feasible set $\mathcal{U}_i$, which is a 2D region, this potentially optimal region $\mathcal{E}_i$ is a part of the circular arc, which is determined only by a one-dimensional range of angles with given circle radius $\bar{d}$. Hence, we can approximately characterize it by a set of discrete points via uniform quantization of its subtending angle, which is able to achieve arbitrarily high accuracy with a sufficiently large number of quantization levels. Let $Q> 1$ denote the number of quantization levels for each intersected coverage boundary. Let $U_{I_i,I_{i+1}}(q)$ represent the $q$th point on the quantized intersected coverage boundary of GBS $I_{i}$ with the coverage area of GBS $I_{i+1}$, where $q\in\mathcal{Q},\mathcal{Q}=\{1,...,Q\}$. The horizontal location of $U_{I_i,I_{i+1}}(q)$ is given by
\begin{align}
{\mv{u}}_{I_i,I_{i+1}}(q)=&\bar{d}\bigg[\cos\left(\phi_{I_i,I_{i+1}}+\left(\frac{q-1}{Q-1}-\frac{1}{2}\right)\theta_{I_i,I_{i+1}}\right),\sin\left(\phi_{I_i,I_{i+1}}+\left(\frac{q-1}{Q-1}-\frac{1}{2}\right)\theta_{I_i,I_{i+1}}\right)\bigg]^T\nonumber\\
&+{\mv{g}}_{I_i},\quad I_i,I_{i+1}\in\mathcal{M},I_i\neq I_{i+1},\
q\in\mathcal{Q},
\end{align}
where $\phi_{I_i,I_{i+1}}=\angle\left([1,j]\times({\mv{g}}_{I_{i+1}}-{\mv{g}}_{I_i})\right)$ denotes the angle between ${\mv{g}}_{I_{i+1}}-{\mv{g}}_{I_i}$ and the x-axis; $\theta_{I_i,I_{i+1}}=2\arccos\left(\|{\mv{g}}_{I_i}-{\mv{g}}_{I_{i+1}}\|/2\bar{d}\right)$ denotes the subtending angle of the arc-shaped intersected coverage boundary, with $\theta_{I_i,I_{i+1}}< \pi$. In Fig. \ref{directed_graph}(a), we illustrate our proposed quantization of the intersected coverage boundaries $\mathcal{E}_i$'s, by taking the example of ${\mv{I}}=[3,1,2,5]^T$ and $Q=2$.

Therefore, an approximate solution to Problem (P3) can be obtained by solving the following problem, which jointly optimizes the GBS-UAV association sequence and the UAV handover locations on the quantized intersected coverage boundaries:
\begin{align}
{\mbox{(P3-II)}}\quad \underset{{\mv{I}},{\mv{q}}}{\min}\ &
\|{\mv{u}}_0-{\mv{u}}_{I_1,I_2}(q_1)\| +\sum_{i=2}^{N-1} \|{\mv{u}}_{I_i,I_{i+1}}(q_i)-{\mv{u}}_{I_{i-1},I_i}(q_{i-1})\|\nonumber\\
&+\|{\mv{u}}_F-{\mv{u}}_{I_{N-1},I_N}(q_{N-1})\|\\
\mathrm{s.t.}\quad & I_i\neq I_j, \quad  \forall i\neq j,\ i,j=1,...,N\label{P6c_different}\\
& q_i\in\mathcal{Q},\quad i=1,...,N-1\label{P6c_Q}\\
& (\ref{feas_i}),(\ref{feas_f}),(\ref{feas_GBS}),(\ref{feas_I}),
\end{align}
where ${\mv{q}}=[q_1,...,q_{N-1}]^T$, with $q_i$ denoting the point index selected from the quantized intersected coverage boundary between the $I_i$th and $I_{i+1}$th GBSs; and the constraints in (\ref{P6c_different}) are resulted from the non-repeated GBS-UAV association property given in Proposition \ref{lemma_I}. It is worth noting that Problem (P3-II) is equivalent to Problem (P3) as $Q\rightarrow\infty$. Although Problem (P3-II) is a non-convex combinatorial optimization problem due to the discrete variables in $\mv{I}$ and $\mv{q}$, we solve it globally optimally by leveraging graph theory in the following.

We start by constructing a directed weighted graph denoted as $G_Q=(V_Q,E_Q)$ based on $U_{I_i,I_{i+1}}(q)$'s introduced above. The vertex set of $G_Q$, denoted as $V_Q$, is given by
\begin{align}\label{V_Q}
V_Q=\{U_0,U_F\}\cup \{U_{m,n}(q):\|{\mv{g}}_m-{\mv{g}}_n\|\leq 2\bar{d},\ m,n\in\mathcal{M}, m\neq n, q\in\mathcal{Q}\}.
\end{align}
The edge set of $G_Q$, denoted as $E_Q$, is given by
\begin{align}\label{E_Q}
&E_Q=\{(U_0,U_{m,n}(q)):\|{\mv{u}}_0-{\mv{g}}_m\|\leq \bar{d},\ m,n\in\mathcal{M},m\neq n,\ q\in\mathcal{Q}\}\nonumber\\
\cup&\{(U_{m,n}(q),U_{n,l}(\hat{q})):\|{\mv{g}}_m-{\mv{g}}_n\|\leq 2\bar{d},\|{\mv{g}}_n-{\mv{g}}_l\|\leq 2\bar{d},\ m,n,l\in\mathcal{M}, m\neq n, n\neq l, m\neq l,\ q,\hat{q}\in\mathcal{Q}\}\nonumber\\
\cup&\{(U_{m,n}(q),U_F):\|{\mv{u}}_F-{\mv{g}}_n\|\leq \bar{d},\ m,n\in\mathcal{M},m\neq n,\ q\in\mathcal{Q}\}.
\end{align}
The weight of each edge is given by
\begin{align}\label{W_Q}
W_Q(U_0,U_{m,n}(q))&=\|{\mv{u}}_0-{\mv{u}}_{m,n}(q)\|,\nonumber\\
W_Q(U_{m,n}(q),U_{n,l}(\hat{q}))&=\|{\mv{u}}_{m,n}(q)-{\mv{u}}_{n,l}(\hat{q})\|,\nonumber\\
W_Q(U_{m,n}(q),U_F)&=\|{\mv{u}}_F-{\mv{u}}_{m,n}(q)\|,\nonumber\\
& m,n\in\mathcal{M}, m\neq n,n\neq l,m\neq l,\ q,\hat{q}\in\mathcal{Q}.
\end{align}
Note that in graph $G_Q$, a vertex $U_{m,n}(q)$ exists if and only if the coverage areas of two different GBSs $m$ and $n$ intersect. Furthermore, an edge $(U_0,U_{m,n}(q))$ or $(U_F,U_{m,n}(q))$ exists if and only if the horizontal location of $U_0$ or $U_F$ lies in the coverage area of GBS $m$ or GBS $n$, respectively, whose weight is given by the distance between $U_0$/$U_F$ and $U_{m,n}(q)$. Moreover, an edge $(U_{m,n}(q),U_{n,l}(\hat{q}))$ exists if and only if the three different GBSs $m$, $n$, and $l$ satisfy the condition that the coverage area of GBS $n$ is intersected with those of both GBS $m$ and GBS $l$, and its weight is given by the distance between $U_{m,n}(q)$ and $U_{n,l}(\hat{q})$. In Fig. \ref{directed_graph}(b), we illustrate the construction of $G_Q$ with the GBS and UAV locations shown in Fig. \ref{directed_graph}(a) and $Q=2$, where the weights are not shown for brevity.

Based on the definition of $G_Q$, finding the optimal ${\mv{I}}$ and ${\mv{q}}$ to Problem (P3-II) is
equivalent to finding the \emph{shortest path} from $U_0$ to $U_F$ in the graph $G_Q$, which can be solved efficiently by e.g., the Dijkstra algorithm \cite{graph}. Note that the maximum number of vertices in $G_Q$ is given by $2+M(M-1)Q$. Thus, the worst-case complexity for constructing graph $G_Q$ and that for finding the shortest path via the Dijkstra algorithm can be shown to be both of $\mathcal{O}(M^4Q^2)$ \cite{graph}.

The above algorithm is summarized as Algorithm \ref{algo_P3_M2} for finding an alternative approximate solution to Problem (P3) with Method II. The worst-case complexity for Algorithm \ref{algo_P3_M2} can be shown to be $\mathcal{O}(M^4Q^2)$ from the analysis above. Next, we evaluate the performance gap between this solution and the optimal solution.
\begin{proposition}\label{prop_complexity}
Denote $\tilde{s}^{\mathrm{II}}$ as the objective value of Problem (P3) with the proposed solution via Method II and $s^\star$ as the optimal value of Problem (P3). We have
\begin{align}\label{bound2}
0\leq\tilde{s}^{\mathrm{II}}-s^\star\leq 4(M-1)\bar{d}\sin\left(\frac{\pi}{4(Q-1)}\right).
\end{align}
\end{proposition}
\begin{IEEEproof}
Please refer to Appendix \ref{proof_prop_complexity}.
\end{IEEEproof}

Note that in general, the performance gap and complexity for Method II decreases and increases as the number of quantization levels, $Q$, increases, respectively. Particularly, the performance gap upper bound shown in (\ref{bound2}) can be well-approximated by $\frac{(M-1)\bar{d}\pi}{Q-1}$ when $Q$ is large, which is inversely proportional to $Q-1$. Hence, Method II is able to achieve a flexible trade-off between performance and complexity by adjusting the value of $Q$. Moreover, it is worth noting that for any given $M$, the performance gap between the proposed solution and the optimal solution approaches zero as $Q$ goes to infinity, while the complexity for Method II is only polynomial in $Q$. This implies that Method II is able to achieve an arbitrarily near-optimal solution by increasing $Q$ at the cost of moderately increased complexity.

\begin{algorithm}[t]\label{algo_P3_M2}
\caption{Proposed Algorithm for Problem (P3) with Method II}
\SetKwData{Index}{Index}
\KwIn{$\bar{d}$, ${\mv{u}}_0$, ${\mv{u}}_F$, $\{{\mv{g}}_m\}_{m=1}^{M}$,
$H$, $H_G$, $V_{\max}$}
\KwOut{$\tilde{\mv{I}}$, $\{\tilde{\mv{u}}^i\}_{i=0}^{\tilde{N}}$}
Construct a graph ${G_Q}=(V_Q,E_Q)$ based on (\ref{V_Q})--(\ref{W_Q}).\\
Find the shortest path from $U_0$ to $U_F$ in ${G}_Q$ via Dijkstra algorithm, and denote the path as $(U_0,U_{\tilde{I}_1,\tilde{I}_2}(q_1^\star(\tilde{\mv{I}})),...,U_{\tilde{I}_{\tilde{N}-1},\tilde{I}_{\tilde{N}}}(q_{\tilde{N}-1}^\star(\tilde{\mv{I}})),U_F)$.\\
Obtain $\tilde{\mv{I}}=[\tilde{I}_1,...,\tilde{I}_{\tilde{N}}]^T$.\\
Obtain $\tilde{\mv{u}}^i={\mv{u}}_{\tilde{I}_i,\tilde{I}_{i+1}}(q_i^\star(\tilde{\mv{I}})),\ i=1,...,\tilde{N}-1$; $\tilde{\mv{u}}^0={\mv{u}}_0$, $\tilde{\mv{u}}^{\tilde{N}}={\mv{u}}_F$.
\end{algorithm}

Finally, in Table \ref{table_P3}, we compare different methods for solving Problem (P3), in terms of complexity and performance gap with the optimal solution.

\begin{table*}[t]
  \centering
  \caption{Comparison of Different Methods for Solving Problem   (P3)}\label{table_P3}
  \renewcommand\arraystretch{1.5}
  \begin{tabular}{|c|c|c|}
  \hline
  Method & Complexity & Performance Gap with Optimal Solution\\
  \hline
  Method I (Exhaustive Search) & $\mathcal{O}(M!M^{3.5})$ & $0$ \\
  \hline
  Method I (Algorithm 1) & $\mathcal{O}(M^{3.5})$ & $\big[0,2M\bar{d}\ \big]$ \\
  \hline
  Method II (Algorithm 2) & $\mathcal{O}(M^4Q^2)$ &   $\big[0,4(M-1)\bar{d}\sin\big(\frac{\pi}{4(Q-1)}\big)\big]$\\
  \hline
  \end{tabular}
\end{table*}

\subsection{Summary of Graph Theory Based Algorithms for Problem (P1)}
In this subsection, we highlight that the proposed two graph theory based algorithms for solving Problem (P1) can be summarized under a unified graph theory based design framework as follows. First, by exploiting the structural properties of (P1) and applying effective bounding and approximation techniques, (P1) is transformed into (P3-I) (with the objective function replaced by its upper bound shown in (\ref{upp})) and (P3-II). Next, two different graphs are constructed for modeling (P3-I) and (P3-II), respectively, under the following general guidelines:
\begin{enumerate}
\item The \emph{existence} of vertices/edges represents the problem (communication connectivity) \emph{constraints}.
\item Each \emph{path} in the graph between a given pair of vertices corresponds to one \emph{feasible solution} to the problem (or a feasible UAV trajectory).
\item The \emph{sum weight} of each path described above equals to the \emph{objective value} of the problem with the corresponding feasible solution.
\end{enumerate}
By this means, (P3-I) and (P3-II) are equivalently transformed into shortest path problems between two given vertices in the correspondingly constructed graphs, which can be efficiently solved via off-the-shelf algorithms in graph theory. Note that the graph connectivity based approach for checking the feasibility of (P1) shown in the preceding section can be similarly shown to also belong to this framework, where the details are omitted for brevity.

\section{Numerical Results}\label{sec_numerical}
In this section, we provide numerical results to evaluate the performance of our proposed trajectory designs. We consider that $M$ GBSs are uniformly randomly distributed in a $D\ \mathrm{km}\times D\ \mathrm{km}$ square region, with density $\lambda=\frac{M}{D^2}\ \mathrm{GBSs}/\mathrm{km}^2$. Specifically, we set $D=10$. The altitude of the UAV and each GBS is set as $H=90$ m and $H_G=12.5$ m, respectively. The maximum UAV speed is set as $V_{\max}=50$ m/s. The reference SNR at distance $d_0=1$ m is set as $\gamma_0=\frac{P\beta_0}{\sigma^2}=80$ dB.

\subsection{Quality-of-Connectivity Performance}
First, we evaluate the quality-of-connectivity performance of the UAV during its mission by our proposed trajectory design. For a given set of GBS locations, let $\bar{\rho}_{\max}$ denote the maximum SNR target that can be achieved for all time instants during the UAV mission. The value of $\bar{\rho}_{\max}$ can be obtained by increasing $\bar{\rho}$ and checking the feasibility of Problem (P1) via the proposed approach in Section \ref{sec_feasibility}, until (P1) becomes infeasible. For comparison, we consider the following simple benchmark trajectory:
\begin{itemize}
\item {\bf{Straight flight (SF) trajectory}}: In this case, the UAV flies from $U_0$ to $U_F$ in a straight line with maximum speed $V_{\max}$.
\end{itemize}
Note that for the SF trajectory, the maximum achievable SNR target is given by
\begin{align}
\bar{\rho}_{\max,\mathrm{SF}}=\frac{\gamma_0}{\underset{0\leq \alpha\leq 1}{\max}\ \underset{m\in\mathcal{M}}{\min}\ \|\alpha {\mv{u}}_0+(1-\alpha){\mv{u}}_F-{\mv{g}}_m\|^2+(H-H_G)^2},
\end{align}
which can be obtained via one-dimensional search over $\alpha$.
\begin{figure}[t]
  \centering
  \includegraphics[width=12cm]{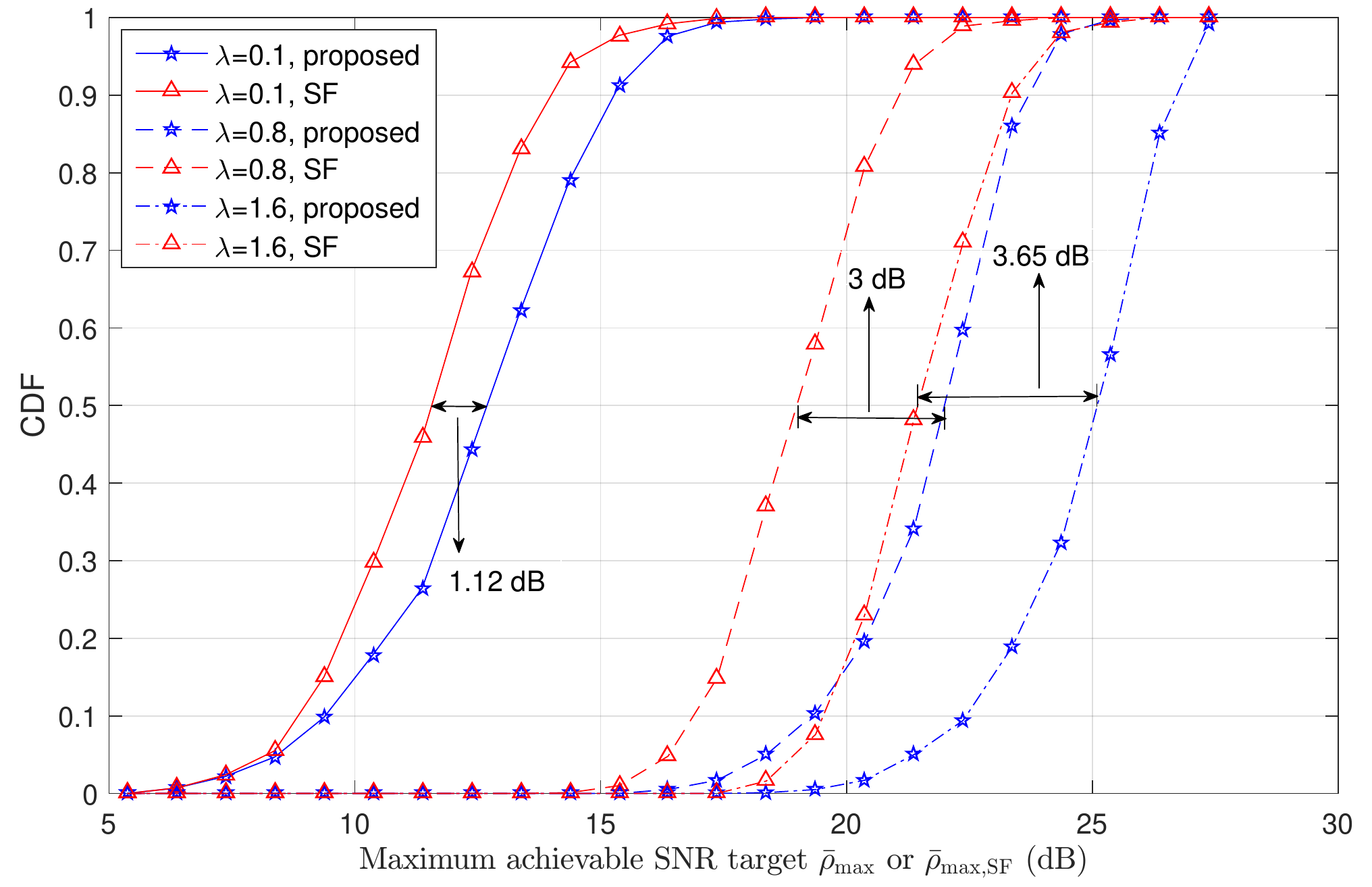}
  \caption{CDF for maximum achievable SNR target $\bar{\rho}_{\max}$ or $\bar{\rho}_{\max,\mathrm{SF}}$ with different GBS density $\lambda$.}\label{CDF}
\end{figure}

We set the UAV's initial and final locations projected on the horizontal plane as ${\mv{u}}_0=[2\ \mathrm{km},2\ \mathrm{km}]^T$ and ${\mv{u}}_F=[8\ \mathrm{km},8\ \mathrm{km}]^T$, respectively. In Fig. \ref{CDF}, we show the cumulative distribution function (CDF) of $\bar{\rho}_{\max}$ for the proposed design and $\bar{\rho}_{\max,\mathrm{SF}}$ for the SF trajectory with different values of the GBS density $\lambda$, where the results for each $\lambda$ are obtained based on $1000$ independent random realizations of the GBS locations. It is observed that as $\lambda$ increases, the SNR performance for both the proposed and the benchmark SF trajectory designs improves, which indicates that the quality-of-connectivity of cellular-enabled UAV communication can be effectively improved by increasing the cellular GBS density. Furthermore, it is observed that the performance gain of our proposed design over the benchmark SF trajectory becomes more significant as $\lambda$ increases. For example, the median value of $\bar{\rho}_{\max}$ that can be achieved for $50\%$ of GBS location realizations with our proposed design is larger than that with the SF trajectory by $1.12$ dB, $3$ dB, and $3.65$ dB, for the case of $\lambda=0.1$, $\lambda=0.8$, and $\lambda=1.6$, respectively. The above results are consistent with our intuition that as the GBS density or the number of GBSs increases in a given area, there is more flexibility for our proposed design to improve the quality-of-connectivity performance, as compared to the heuristic SF trajectory.

\subsection{Mission Completion Time}
\begin{figure}[!htb]
  \centering
  \subfigure[$\bar{\rho}=17$ dB]{
  \includegraphics[width=8cm]{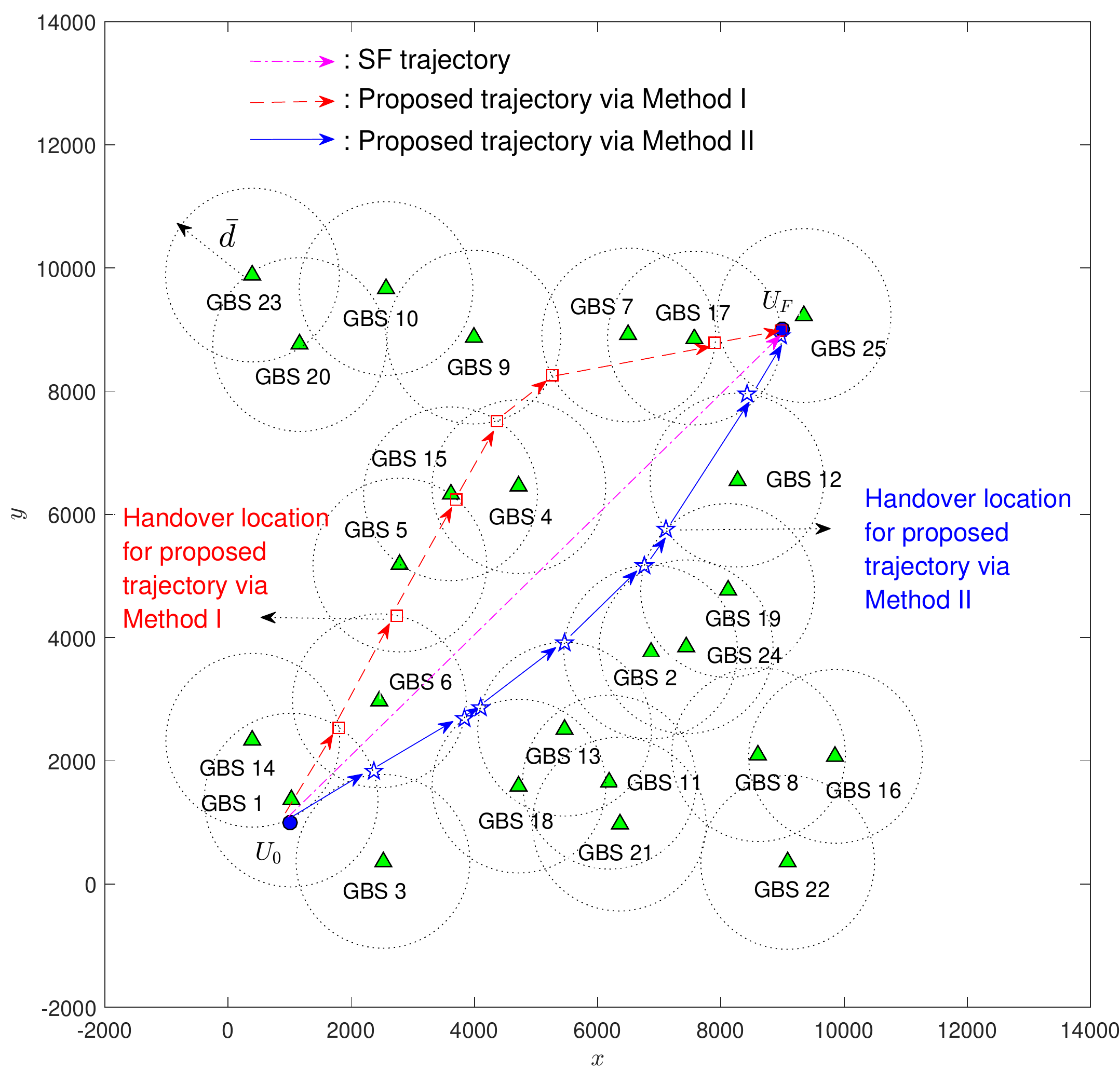}}
  \subfigure[$\bar{\rho}=17.97$ dB]{
  \includegraphics[width=8cm]{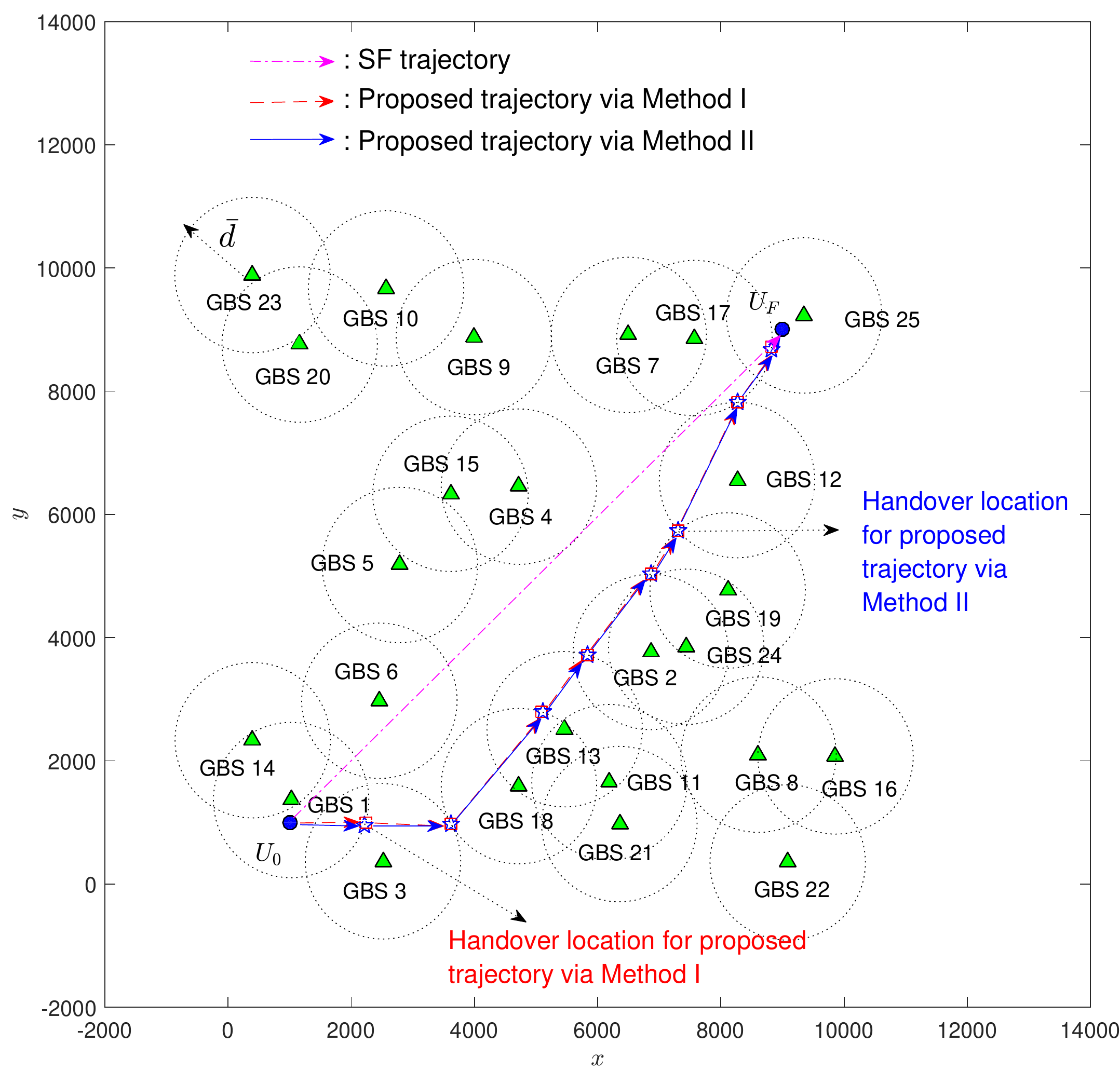}}
  \caption{Illustration of different trajectory designs.}\label{trajectory}
\end{figure}

Next, we evaluate the mission completion time required to achieve a given quality-of-connectivity or SNR target for our proposed trajectory designs given in Section \ref{sec_solution} via Method I or Method II. For this simulation, we randomly generate the locations of $M=25$ GBSs in a $10\ \mathrm{km}\times 10\ \mathrm{km}$ square region, as shown in Fig. \ref{trajectory}. The UAV's initial and final locations projected on the horizontal plane are set as ${\mv{u}}_0=[1\ \mathrm{km},1\ \mathrm{km}]^T$ and ${\mv{u}}_F=[9\ \mathrm{km},9\ \mathrm{km}]^T$, respectively. For comparison, we also consider the SF trajectory as a benchmark. In addition, we consider another benchmark trajectory design:
\begin{itemize}
\item {\bf{Optimal trajectory}}: In this case, the UAV trajectory follows the optimal solution to Problem (P1), which is obtained by finding the optimal solution to Problem (P3) through Method I via exhaustive search over all feasible GBS-UAV associations.\footnote{This is feasible for the considered setup with a moderate value of $M$, while the complexity goes up exponentially with $M$ and thus becomes infeasible for practical setups with much larger $M$.}
\end{itemize}

\begin{figure}[!htb]
  \centering
  \includegraphics[width=12cm]{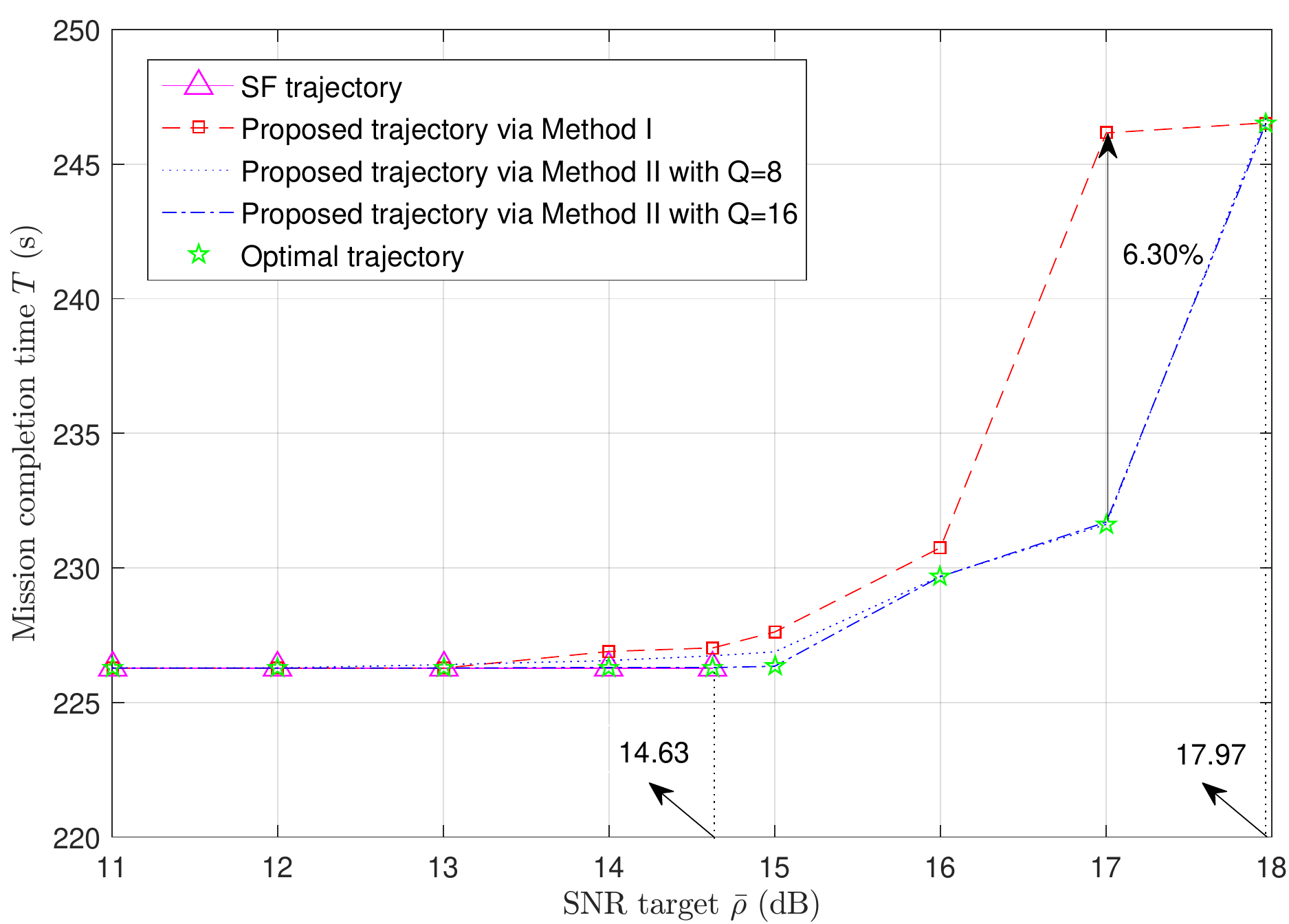}
  \caption{Mission completion time ${T}$ versus SNR target $\bar{\rho}$ with different trajectory designs.}\label{time}
\end{figure}

In Fig. \ref{time}, we show the mission completion time $T$ versus the SNR target $\bar{\rho}$ with our proposed trajectory designs via Method I or Method II, the SF trajectory, and the optimal trajectory. For the proposed design with Method II, we set $Q=8$ or $Q=16$. It is observed that although the SF trajectory achieves minimum mission completion time, it becomes infeasible as the SNR target exceeds $\bar{\rho}_{\max,\mathrm{SF}}=14.63$ dB, which is $3.34$ dB smaller compared to the maximum achievable SNR target for other trajectory designs given by $\bar{\rho}_{\max}=17.97$ dB. Moreover, it is observed that our proposed trajectory design via Method II for both cases of $Q=8$ and $Q=16$ achieves almost the same performance as the optimal trajectory, which thus validates the near-optimality of this design as shown in Section \ref{sec_P3}. In addition, it is observed that the performance of this design is improved by increasing the number of quantization levels from $Q=8$ to $Q=16$, which demonstrates its flexibility for performance-complexity trade-off as discussed in Section \ref{sec_P3}. Furthermore, the proposed trajectory design via Method I generally performs closely to the optimal trajectory, but yields $6.30\%$ more mission completion time at the SNR target of $\bar{\rho}=17$ dB. This is because the upper bound used for approximating the objective function of Problem (P3-I) may not be always tight, as explained in Section \ref{sec_P3}.

Finally, we fix two SNR targets as $\bar{\rho}=17$ dB and $\bar{\rho}=17.97$ dB, and illustrate their corresponding trajectory designs in Fig. \ref{trajectory}(a) and Fig. \ref{trajectory}(b), respectively. Specifically, we show the trajectory designs via Method I, Method II with $Q=16$, as well as the SF trajectory. The optimal trajectory is not illustrated for brevity since it is observed to yield the same GBS-UAV association sequence and almost the same handover locations as the trajectory obtained via Method II. For the case of $\bar{\rho}=17$ dB, it is observed from Fig. \ref{trajectory}(a) that the proposed trajectory design via Method I is quite different from that via Method II, since Method I selects a suboptimal GBS-UAV association sequence which substantially deviates from the optimal one selected by Method II. This further explains the performance loss of Method I compared to Method II as mentioned above. On the other hand, for the case of $\bar{\rho}=17.97$ dB, it is observed from Fig. \ref{trajectory}(b) that the proposed trajectory designs via Method I and Method II are almost the same, which is consistent with their similar performances as shown in Fig. \ref{time}. Last, it can be observed from Fig. \ref{trajectory}(a) and Fig. \ref{trajectory}(b) that the SF trajectory cannot reach the SNR targets for all time in both cases.

To summarize, our proposed trajectory designs via Method I and Method II in general achieve near-optimal performance and significantly outperform the simple SF trajectory. In addition, the trajectory design solution is critically dependent on the given quality-of-connectivity requirement. This thus validates the importance of trajectory optimization under the new connectivity constraint studied in this paper.

\section{Concluding Remarks and Future Work}\label{sec_conclusion}
In this paper, we first provided an overview on the two important new paradigms in cellular UAV communications, namely, cellular-enabled UAV communication and UAV-assisted cellular communication. Then, we focused on the design of a cellular-enabled UAV communication system from a connectivity-constrained trajectory optimization perspective, where a UAV has a mission of flying from an initial location to a final location while maintaining a target quality-of-connectivity with the cellular network. Specifically, we formulated the UAV trajectory optimization problem to minimize the mission completion time, subject to a minimum received SNR constraint of the UAV-cellular communication link, the UAV's initial and final location constraints, as well as its maximum speed constraint. We established a graph theory based design framework for checking the feasibility and finding high-quality approximate solutions of the formulated trajectory optimization problem. Analytical results showed that the devised algorithms are able to achieve a flexible trade-off between complexity and performance, and approach the optimal solution with an arbitrarily small performance gap with only polynomial complexity. Numerical results further validated the efficacy of our proposed trajectory designs.

The proposed trajectory design approaches and methodologies in this paper can be applied/
extended to solve a class of communication-aware UAV trajectory optimization problems under more general setups and with other performance considerations, some of which are discussed as follows to motivate future work.
\begin{itemize}
\item \emph{3D trajectory optimization with altitude mask constraint:} Generally speaking, the quality-of-connectivity of the GBS-UAV communication link improves as the UAV altitude decreases under our considered LoS channel model, due to the reduced distance-dependent path-loss. Hence, it is desirable to let the UAV fly at its minimum allowable altitude under safety consideration. In this paper, we assume a constant minimum altitude $H$ of the UAV, thus the UAV trajectory only needs to be designed in a 2D horizontal plane. However, in practice, the minimum allowable altitude may change over locations due to different terrains (buildings, hills, trees, etc.). Moreover, a maximum allowable UAV altitude may also be specified in many countries. As a result, in general we can model the UAV altitude constraint as a location-dependent \emph{``altitude mask''}, which specifies the maximum and minimum allowable altitude at each horizontal location. Note that extending our proposed 2D trajectory design in this paper to the 3D design under the above new altitude mask constraint is a non-trivial task, since it may not be optimal to always let the UAV fly at its minimum allowable altitudes, as other factors such as the altitudes of the initial and final locations, the UAV maximum speed, as well as the altitude mask spatial distribution need to be considered.
\item \emph{Trajectory optimization with CoMP-based cellular-enabled UAV communication:} In this paper, we assume that the UAV is associated with one GBS that has the best channel with the UAV at each time instant. In practice, to further enhance the link connectivity, multiple GBSs can be cooperatively associated with the UAV at each time instant by leveraging the LoS-induced macro-diversity and the \emph{CoMP} transmission/reception. For example, consider the uplink transmission where a UAV can be simultaneously associated with $1<\tilde{M}(t)\leq M$ GBSs at each time $t$, represented by a time-varying set $\tilde{\mathcal{M}}(t)\subseteq\mathcal{M}$, which cooperatively perform maximum ratio combining (MRC) for detection. The SNR at the combining output is then given by
    \begin{align}
    \rho(t)=\sum_{m\in\tilde{\mathcal{M}}(t)}\frac{\gamma_0}{(H-H_G)^2+\|{\mv{u}}(t)-g_m\|^2},\quad
    0\leq t\leq T,
    \end{align}
    where $\gamma_0$ is similarly defined as in Section \ref{sec_system}. In this case, the sets of GBSs that are sequentially associated with the UAV during its mission, i.e.,     $\{\tilde{\mathcal{M}}(t),\ 0\leq t\leq T\}$, need to be jointly optimized with the UAV trajectory $\{{\mv{u}}(t),\ 0\leq t\leq T\}$, which is a more challenging problem compared to the case of $\tilde{M}(t)=1$ and $\tilde{\mathcal{M}}(t)=\left\{\arg\underset{m\in\mathcal{M}}{\min}\ {\|{\mv{u}}(t)-{\mv{g}}_m\|}\right\}$ as considered in this paper, due to the more complicated coupling between $\{\tilde{\mathcal{M}}(t),\ 0\leq t\leq T\}$ and the trajectory $\{{\mv{u}}(t),\ 0\leq t\leq T\}$.
\item \emph{Trajectory optimization with other QoS requirements in cellular-enabled UAV communication}: In this paper, we consider the UAV trajectory design under the quality-of-connectivity constraint of the GBS-UAV communication link, which is suitable for delay-sensitive communication applications; while other QoS constraints may apply for   different scenarios in cellular-enabled UAV communication. For example, for delay-limited GBS-UAV communication, another QoS metric is the so-called \emph{connectivity outage}, which is defined as the time duration of the quality-of-connectivity not meeting a certain threshold over the UAV mission. Moreover, for the uplink rate-demanding UAV payload data transmission that can tolerate certain delay, a relevant performance metric is the \emph{average achievable rate} over the UAV mission. How to extend our results to the UAV trajectory designs considering the above other QoS requirements is an interesting direction worthy of further investigation.
\item \emph{Online UAV trajectory optimization under non-LoS channel model}: In this paper, we adopt the \emph{LoS channel model} between the GBSs and the UAV, which is a suitable choice when GBSs are located in rural or suburban environments, or when the UAV altitude is sufficiently high. In this case, the channel power gain between each GBS and the UAV at any location is solely dependent on their distance as shown in (\ref{channel}), which entails the \emph{offline} UAV trajectory design approach, as considered in this paper. On the other hand, when GBSs are located in urban environments or the UAV flies at a moderate or low altitude, other non-LoS channel models may apply due to the existence of possible scatters/blockages, such as the \emph{probabilistic LoS channel model} where the LoS probability in general increases with the altitude of the UAV \cite{ModelingA2G,LAP}, and the \emph{Rician fading channel model} with both LoS and scattered fading components \cite{MChannel}. In this case, it is generally impossible to obtain the complete channel knowledge between each GBS and every possible UAV location prior to the UAV mission. As a result, a combined online and offline trajectory design is more promising in this case, where an offline trajectory is first designed before the mission based on the partial/statistical channel knowledge and the trajectory is then adjusted in real-time (or online) during the mission by the UAV based on its measured/received signals from the GBSs along its trajectory.
\item \emph{Trajectory optimization with interference consideration}: In this paper, we assume that the considered UAV is allocated with an exclusive frequency band, thus the GBS-UAV communication link is interference-free. In practice, the frequency band may be reused by other UAVs and/or ground users to improve the spectral efficiency, and as a result the interference issue needs to be taken into account in the trajectory design. For instance, for scenarios where multiple UAVs need to share the same band, the inter-UAV interference could be mitigated by jointly designing their trajectories to ensure maximum geographical separations at each time instant; while for some UAVs that cannot be sufficiently separated over time, communication resource allocation such as dynamic channel assignment and power control can be employed to mitigate the co-channel interference \cite{JointTC}.
\item \emph{Trajectory optimization with UAV energy consumption consideration}: As discussed in Section \ref{sec_intro}, the limited on-board energy and the high propulsion energy consumption of UAVs are important issues in the UAV trajectory design, which, however, have not been considered in this paper. By taking the practical UAV energy considerations into account, Problem (P1) can be modified to minimize the mission completion time subject to additional constraints on the propulsion power/energy consumptions of the UAV along its trajectory. Such new problems need accurate energy consumption models for different types of UAV \cite{energy,Rotary} and are expected to reveal interesting new trade-offs between the communication performance and energy consumption in the UAV trajectory design.
\end{itemize}

\appendices
\section{Proof of Proposition \ref{prop_feas}}\label{proof_prop_feas}
First, we prove the ``if'' part by showing that a feasible solution to Problem (P1) can be found with any given GBS-UAV association sequence $\mv{I}$ that satisfies the conditions in (\ref{feas_i})--(\ref{feas_I}). Specifically, we let the UAV fly from $U_0$ to $U_F$ following a path consisting of $N$ connected line segments specified by $N+1$ discrete points with same altitude $H$, where the horizontal locations of the starting and ending points of each $i$th line segment are denoted as ${\mv{u}}^{i-1}$ and ${\mv{u}}^i$, respectively. The set $\{{\mv{u}}^i\}_{i=0}^N$ is given by ${\mv{u}}^0={\mv{u}}_0$, ${\mv{u}}^N={\mv{u}}_F$, and
\begin{align}
{\mv{u}}^i={\mv{g}}_{I_i}+\bar{d}\frac{{\mv{g}}_{I_{i+1}}-{\mv{g}}_{I_i}}{\|{\mv{g}}_{I_{i+1}}-{\mv{g}}_{I_i}\|},\quad
i=1,...,N-1.
\end{align}
It can be shown from (\ref{feas_i}), (\ref{feas_f}) and (\ref{feas_GBS}) that
\begin{align}
&\|{\mv{u}}^i-{\mv{g}}_{I_i}\|\leq \bar{d},\quad i=1,...,N\label{if1}\\
&\|{\mv{u}}^i-{\mv{g}}_{I_{i+1}}\|\leq \bar{d},\quad i=0,...,N-1.\label{if2}
\end{align}
Therefore, for any point in the $i$th line segment with horizontal location ${\mv{u}}^i(p)=p{\mv{u}}^{i-1}+(1-p){\mv{u}}^{i},\ \forall p\in [0,1]$, we have
\begin{align}\label{u3}
&\|{\mv{u}}^i(p)-{\mv{g}}_{I_i}\|=\|p({\mv{u}}^{i-1}-{\mv{g}}_{I_i})+(1-p)({\mv{u}}^{i}-{\mv{g}}_{I_i})\|\nonumber\\
\!\!\overset{(A_1)}{\leq}\!&p\|{\mv{u}}^{i-1}\!-\!{\mv{g}}_{I_i}\|\!+\!(1-p)\|{\mv{u}}^{i}\!-\!{\mv{g}}_{I_i}\|\!\!\overset{(A_2)}{\leq}\!\!\bar{d},\quad i=1,...,N,
\end{align}
where $(A_1)$ is due to the triangle inequality, and $(A_2)$ is resulted from (\ref{if1}) and (\ref{if2}).

It then follows from (\ref{u3}) that with the above UAV path and arbitrary UAV speed that satisfies $\|\dot{\mv{u}}(t)\|\leq V_{\max}$, the resulting UAV trajectory satisfies (\ref{P1c_i})--(\ref{P1c_v}), which thus completes the proof of the ``if'' part.

Next, we prove the ``only if'' part by showing that given any feasible solution $(T,\{{\mv{u}}(t),\ 0\leq t\leq T\})$ to Problem (P1), we can always construct $\{I_i\}_{i=1}^N$ that satisfies the conditions in (\ref{feas_i})--(\ref{feas_I}). Specifically, we can always find a finite number $N$ to divide $[0,T]$ into $N$ intervals and construct $\{I_i\}_{i=1}^N$, where $\arg\underset{m\in \mathcal{M}}{\min}\ \|{\mv{u}}(t)-{\mv{g}}_m\|=I_i$ and $\|{\mv{u}}(t)-{\mv{g}}_{I_i}\|\leq \bar{d}$ hold when $t$ lies in the $i$th interval, $\forall i\in\{1,...,N\}$. Note that the conditions in (\ref{feas_I}) are automatically satisfied by $\{I_i\}_{i=1}^N$. Then, we construct $\{{\mv{u}}^i\}_{i=0}^N$ by defining ${\mv{u}}^0={\mv{u}}_0$ and letting ${\mv{u}}^i$ denote the horizontal location of the UAV at the end of the $i$th interval, $\forall i\in \{1,...,N\}$, with ${\mv{u}}^N={\mv{u}}_F$. It follows that the constructed $\{I_i\}_{i=1}^N$ and $\{{\mv{u}}^i\}_{i=0}^N$ satisfy (\ref{if1}) and (\ref{if2}). As a result, it can be readily shown that $\{I_i\}_{i=1}^N$ satisfies the conditions in (\ref{feas_i}) and (\ref{feas_f}). Moreover, we have
\begin{align}
&\|{\mv{g}}_{I_{i+1}}-{\mv{g}}_{I_i}\|=\|({\mv{u}}^i-{\mv{g}}_{I_i})-({\mv{u}}^i-{\mv{g}}_{I_{i+1}})\|\nonumber\\
\!\!\overset{(A_3)}{\leq}\!&\|{\mv{u}}^i-{\mv{g}}_{I_i}\|\!+\!\|{\mv{u}}^i-{\mv{g}}_{I_{i+1}}\|\!\overset{(A_4)}{\leq}\!
2\bar{d},\quad  i=1,...,N-1,\label{onlyif3}
\end{align}
where $(A_3)$ is due to the triangle inequality, and $(A_4)$ results from (\ref{if1}) and (\ref{if2}). Hence, the constructed $\{I_i\}_{i=1}^N$ also satisfies (\ref{feas_GBS}), which thus completes the proof of the ``only if'' part. By combining the proofs of both the ``if'' and ``only if'' parts, the proof of Proposition \ref{prop_feas} is completed.

\section{Proof of Proposition \ref{prop_P1eq}}\label{proof_prop_P1eq}
First, given any feasible solution $(T,\{{\mv{u}}(t),\ 0\leq t\leq T\},{\mv{I}},\{T_i\}_{i=1}^N)$ to Problem (P2), it follows from (\ref{P2c_SNR}) that $\underset{m\in\mathcal{M}}{\min}\ \|{\mv{u}}(t)-{\mv{g}}_m\|\leq \|{\mv{u}}(t)-{\mv{g}}_{I_i}\|\leq \bar{d}$ holds for any $t\in\left[\sum_{j=1}^{i-1}T_j,\sum_{j=1}^i T_j\right]$ and $i=1,...,N$. Thus, $(T,\{{\mv{u}}(t),\ 0\leq t\leq T\})$ is a feasible solution to Problem (P1) and achieves the same objective value as Problem (P2) with the solution $(T,\{{\mv{u}}(t),\ 0\leq t\leq T\},{\mv{I}},\{T_i\}_{i=1}^N)$. Hence, the optimal value of Problem (P1) is no larger than that of Problem (P2). On the other hand, for any given feasible solution $(T,\{{\mv{u}}(t),\ 0\leq t\leq T\})$ to Problem (P1), we can always divide $[0,T]$ into $N$ intervals denoted by $\left[\sum_{j=1}^{i-1}T_j, \sum_{j=1}^i T_j\right],\ i=1,...,N$, such that $I_i=\arg\underset{m\in\mathcal{M}}{\min}\ \|{\mv{u}}(t)-{\mv{g}}_m\|$ and $\|{\mv{u}}(t)-{\mv{g}}_{I_i}\|\leq \bar{d}$ hold for any $t\in\left[\sum_{j=1}^{i-1}T_j,\sum_{j=1}^i T_j\right]$ and $i=1,...,N$. By following similar procedure as in the ``only if'' part of the proof of Proposition \ref{prop_feas}, it can be shown that ${\mv{I}}=[I_1,...,I_N]^T$ satisfies the constraints in (\ref{feas_i})--(\ref{feas_I}). Hence, $(T,\{{\mv{u}}(t),\ 0\leq t\leq T\},{\mv{I}},\{T_i\}_{i=1}^N)$ can be shown to be feasible for Problem (P2) and achieves the same objective value as Problem (P1) with the solution $(T,\{{\mv{u}}(t),\ 0\leq t\leq T\})$. The optimal value of Problem (P2) is thus no larger than that of Problem (P1). Therefore, Problem (P1) and Problem (P2) have the same optimal value, which completes {\hbox{the proof of Proposition \ref{prop_P1eq}.}}

\section{Proof of Proposition \ref{prop_P2}}\label{proof_prop_P2}
We prove Proposition \ref{prop_P2} by showing that for any feasible solution to Problem (P2) denoted by $(\tilde{T},\{\tilde{\mv{u}}(t),\ 0\leq t\leq \tilde{T}\},{\mv{I}},\{\tilde{T}_i\}_{i=1}^N)$, we can always construct a feasible solution to Problem (P2) denoted by $({T},\{{\mv{u}}(t),\ 0\leq t\leq T\},{\mv{I}},\{T_i\}_{i=1}^N)$ that satisfies the conditions in (\ref{Ti}), (\ref{ut}) and (\ref{T}), and achieves no larger objective value of Problem (P2) compared to $(\tilde{T},\{\tilde{\mv{u}}(t),\ 0\leq t\leq \tilde{T}\},{\mv{I}},\{\tilde{T}_i\}_{i=1}^N)$. We start by constructing the same handover locations in $\{{\mv{u}}(t),\ 0\leq t\leq T\}$ as those in $\{\tilde{\mv{u}}(t),\ 0\leq t\leq \tilde{T}\}$, i.e., ${\mv{u}}^i=\tilde{\mv{u}}\left(\sum_{j=1}^i \tilde{T}_j\right),\ i=0,...,N$. Then, note that $\tilde{T}_i$ denotes the time duration for the UAV to fly from ${\mv{u}}^{i-1}$ to ${\mv{u}}^i$, thus $\tilde{T}_i\geq \frac{\|{\mv{u}}^i-{\mv{u}}^{i-1}\|}{V_{\max}},\ i=1,...,N$ should hold, since $\|{\mv{u}}^i-{\mv{u}}^{i-1}\|$ is the minimum distance between ${\mv{u}}^{i-1}$ and ${\mv{u}}^i$, and $V_{\max}$ is the maximum allowable speed. By noting that $T_i=\frac{\|{\mv{u}}^i-{\mv{u}}^{i-1}\|}{V_{\max}}$ holds as shown in (\ref{Ti}), we have $T_i\leq \tilde{T}_i,\ i=1,...,N$, and consequently $\tilde{T}=\sum_{i=1}^N\tilde{T}_i\geq T=\sum_{i=1}^N{T}_i$. The proof of Proposition \ref{prop_P2} is thus completed.

\section{Proof of Proposition \ref{lemma_I}}\label{proof_lemma_I}
Consider a feasible solution of ${\mv{I}}$ to Problem (P3) given by $\hat{\mv{I}}=\left[\hat{I}_1,...,\hat{I}_k,...,\hat{I}_q,...,I_{\hat{N}}\right]^T$, where $\hat{I}_k=\hat{I}_q$, and another feasible solution of $\mv{I}$ by removing the $(k+1)$th to the $q$th elements in $\hat{\mv{I}}$, which is given by $\tilde{\mv{I}}=\left[\hat{I}_1,...,\hat{I}_k,\hat{I}_{q+1},...,\hat{I}_{\hat{N}}\right]^T$. It can be shown that for given $\hat{\mv{I}}$ and any feasible $\{\hat{\mv{u}}^i\}_{i=0}^{\hat{N}}$ to Problem (P3), the resulted objective value is given by $\hat{s}\overset{\Delta}{=}\sum_{i=1}^{k-1} \|\hat{\mv{u}}^i-\hat{\mv{u}}^{i-1}\|+\sum_{i=k}^q \|\hat{\mv{u}}^i-\hat{\mv{u}}^{i-1}\|+\sum_{i=q+1}^{\hat{N}}\|\hat{\mv{u}}^i-\hat{\mv{u}}^{i-1}\|$. On the other hand, it can be shown that $(\tilde{\mv{I}},\{\tilde{\mv{u}}^i\}_{i=0}^{\hat{N}-(q-k)})$ with $\tilde{\mv{u}}^i=\hat{\mv{u}}^i,\ i=0,...,k-1$ and $\tilde{\mv{u}}^i=\hat{\mv{u}}^{i+(q-k)},\ i=k,...,\hat{N}-(q-k)$ is also a feasible solution to Problem (P3), whose objective value is given by $\tilde{s}\overset{\Delta}{=}\sum_{i=1}^{k-1} \|\hat{\mv{u}}^i-\hat{\mv{u}}^{i-1}\|+\|\hat{\mv{u}}^q-\hat{\mv{u}}^{k-1}\|+\sum_{i=q+1}^{\hat{N}}\|\hat{\mv{u}}^i-\hat{\mv{u}}^{i-1}\|$. By applying the triangle inequality, it can be shown that $\|\hat{\mv{u}}^q-\hat{\mv{u}}^{k-1}\|=\|\sum_{i=k}^q(\hat{\mv{u}}^i-\hat{\mv{u}}^{i-1})\|\leq \sum_{i=k}^q\|\hat{\mv{u}}^i-\hat{\mv{u}}^{i-1}\|$ holds. It then follows that $\tilde{s}\leq \hat{s}$ holds, i.e., the objective value of Problem (P3) with the solution $(\tilde{\mv{I}},\{\tilde{\mv{u}}^i\}_{i=0}^{\hat{N}-(q-k)})$ is no larger than that of Problem (P3) with the solution $(\hat{\mv{I}},\{\hat{\mv{u}}^i\}_{i=0}^{\hat{N}})$. Therefore, the optimal solution to Problem (P3) should satisfy $I_i\neq I_j,\ \forall i\neq j$, and thus the length of the optimal $\mv{I}$ should not exceed the total number of GBSs, i.e., $N\leq M$ should hold. This completes the proof of Proposition \ref{lemma_I}.

\section{Proof of Proposition \ref{prop_edge}}\label{proof_prop_edge}
Suppose that the optimal solution to Problem (P3) denoted by $(\{{\mv{u}}^i\}_{i=0}^N,{\mv{I}})$ does not satisfy (\ref{edge}) and (\ref{Ei}), we show that an alternative solution denoted by $(\{{{\mv{u}}^i}^\star\}_{i=0}^N,{\mv{I}}^\star)$ with ${\mv{I}}^\star={\mv{I}}$ can always be found which satisfies (\ref{edge}) and (\ref{Ei}), and achieves no larger objective value of Problem (P3) as compared to $(\{{\mv{u}}^i\}_{i=0}^N,{\mv{I}})$. To start with, we provide the following lemma.
\begin{lemma}\label{lemma_alpha}
For a feasible solution $(\{{\mv{u}}^i\}_{i=0}^N,{\mv{I}})$ to Problem (P3) to be optimal, there always exists $\{\alpha_i\}_{i=1}^{N-1}$ that satisfies the following conditions:
\begin{align}
\|{{\mv{u}}^i}+\alpha_i({\mv{u}}^{i+1}-{\mv{u}}^i)-{\mv{g}}_{{I}_i}\|=\bar{d},\quad i=1,...,N-1&\label{alpha_1}\\
\alpha_i\in [0,1],\quad i=1,...,N-1&.\label{alpha_2}
\end{align}
\end{lemma}
\begin{IEEEproof}
Notice that (\ref{alpha_1}) can be equivalently rewritten as $f_i(\alpha_i)=\|{\mv{u}}^i+\alpha_i({\mv{u}}^{i+1}-{\mv{u}}^i)-{\mv{g}}_{{I}_i}\|^2=\alpha_i^2\|{\mv{u}}^{i+1}-{\mv{u}}^i\|^2+2\alpha_i({\mv{u}}^i-{\mv{g}}_{{I}_i})^T({\mv{u}}^{i+1}-{\mv{u}}^i)+\|{\mv{u}}^i-{\mv{g}}_{{I}_i}\|^2=\bar{d}^2,\ i=1,...,N-1$. Based on this, it can be shown that there always exists $\alpha_i\in [0,1]$ that satisfies (\ref{alpha_1}) if $f_i(0)\leq \bar{d}^2$ and $f_i(1)\geq\bar{d}^2$ hold. In the following, we prove Lemma \ref{lemma_alpha} by showing that with the optimal solution to (P3), $f_i(0)\leq \bar{d}^2$ and $f_i(1)\geq\bar{d}^2$ hold for $i=1,...,N-1$.

First, it follows from (\ref{P3c_SNR1}) directly that $f_i(0)=\|{\mv{u}}^i-{\mv{g}}_{ {I}_i}\|^2\leq \bar{d}^2,\ \forall i$ holds. On the other hand, we show that $f_i(1)=\|{\mv{u}}^{i+1}-{\mv{g}}_{ {I}_i}\|^2\geq \bar{d}^2,\ \forall i$ holds by contradiction. Suppose $(\{\tilde{\mv{u}}^i\}_{i=0}^{\tilde{N}},\tilde{\mv{I}})$ is the optimal solution to Problem (P3) with $f_{\tilde{i}}(1)< \bar{d}^2$, $\tilde{i}\in\{1,...,\tilde{N}-1\}$, we show that a new feasible solution to Problem (P3) denoted as $(\{{\mv{u}}^i\}_{i=0}^N,{\mv{I}})$ can be always constructed, which achieves smaller objective value of Problem (P3) compared to $(\{\tilde{\mv{u}}^i\}_{i=0}^{\tilde{N}},\tilde{\mv{I}})$. Specifically, we set $N=\tilde{N}-1$; ${\mv{u}}^i=\tilde{\mv{u}}^i$ for $i=0,...,\tilde{i}-1$, ${\mv{u}}^i=\tilde{\mv{u}}^{i+1}$ for $i=\tilde{i},...,N$; $I_i=\tilde{I}_i$ for $i=1,...,\tilde{i}$, and $I_i=\tilde{I}_{i+1}$ for $i=\tilde{i}+1,...,N$. It can be shown that $(\{{\mv{u}}^i\}_{i=0}^N,{\mv{I}})$ satisfies all constraints in Problem (P3); moreover, we have $\sum_{i=1}^N\|{\mv{u}}^i-{\mv{u}}^{i-1}\|=\sum_{i=1}^{\tilde{i}-1} \|\tilde{\mv{u}}^i-\tilde{\mv{u}}^{i-1}\|+\|\tilde{\mv{u}}^{\tilde{i}+1}-\tilde{\mv{u}}^{\tilde{i}-1}\|+\sum_{i=\tilde{i}+2}^{\tilde{N}}\|\tilde{\mv{u}}^{i}-\tilde{\mv{u}}^{i-1}\|\overset{(E_1)}{<} \sum_{i=1}^{\tilde{N}}\|\tilde{\mv{u}}^i-\tilde{\mv{u}}^{i-1}\|$, where $(E_1)$ is due to the triangle inequality. This thus completes the proof of Lemma \ref{lemma_alpha}.
\end{IEEEproof}

Then, based on Lemma \ref{lemma_alpha}, we proceed to introduce the construction of such $(\{{{\mv{u}}^i}^\star\}_{i=0}^N,{\mv{I}}^\star)$ according to $(\{{{\mv{u}}^i}\}_{i=0}^N,{\mv{I}})$. Specifically, we construct $\{{{\mv{u}}^i}^\star\}_{i=0}^N$ by setting ${{\mv{u}}^0}^\star={\mv{u}}_0$, ${{\mv{u}}^N}^\star={\mv{u}}_F$, and
\begin{align}
{{\mv{u}}^i}^\star={\mv{u}}^i+\alpha_i({{\mv{u}}^{i+1}}-{{\mv{u}}^i}),\quad i=1,...,N-1,
\end{align}
where $\alpha_i$ satisfies (\ref{alpha_1}) and (\ref{alpha_2}). It can be easily verified that $(\{{{\mv{u}}^i}^\star\}_{i=0}^N,{\mv{I}}^\star)$ is feasible for Problem (P3). Moreover, we have
\begin{align}
&\sum_{i=1}^N\|{{\mv{u}}^i}^\star-{{\mv{u}}^{i-1}}^\star\|=\sum_{i=1}^N\|({{\mv{u}}^i}^\star-{\mv{u}}^i)+({\mv{u}}^i-{{\mv{u}}^{i-1}}^\star)\| \leq\sum_{i=1}^N (\|{{\mv{u}}^i}^\star-{\mv{u}}^i\|+\|{\mv{u}}^i-{{\mv{u}}^{i-1}}^\star\|)\nonumber\\
=&\sum_{i=1}^{N-1}\alpha_i\|{\mv{u}}^{i+1}-{\mv{u}}^i\|+\sum_{i=2}^{N}(1-\alpha_{i-1})\|{\mv{u}}^{i}-{\mv{u}}^{i-1}\|+\|{\mv{u}}^1-{\mv{u}}^0\|=\sum_{i=1}^N\|{\mv{u}}^i-{\mv{u}}^{i-1}\|.
\end{align}
This thus completes the proof of Proposition \ref{prop_edge}.

\section{Proof of Proposition \ref{prop_complexity}}\label{proof_prop_complexity}
Let $({\mv{I}}^\star,\{{{\mv{u}}^i}^\star\}_{i=0}^{N^\star})$ denote the optimal solution to Problem (P3) and $(\tilde{\mv{I}},{\mv{q}}^\star(\tilde{\mv{I}}))$ denote the optimal solution to Problem (P3-II). Define $\hat{q}_i({\mv{I}}^\star)=\arg\underset{q\in\mathcal{Q}}{\min}\ \|{{\mv{u}}^i}^\star-{\mv{u}}_{{I}_i^\star,{I}_{i+1}^\star}(q)\|$ as the index for the point on the quantized intersected coverage boundary between GBSs ${I}_i^\star$ and ${I}_{i+1}^\star$ that is closest to ${{\mv{u}}^i}^\star$. Based on the above definition, we have
\begin{align}\label{boundII_1}
\tilde{s}^{\mathrm{II}}=&\|{\mv{u}}_0-{\mv{u}}_{\tilde{I}_1,\tilde{I}_2}(q_1^\star(\tilde{\mv{I}}))\|+\sum_{i=2}^{\tilde{N}-1}\|{\mv{u}}_{\tilde{I}_{i},\tilde{I}_{i+1}}(q_{i}^\star(\tilde{\mv{I}}))-{\mv{u}}_{\tilde{I}_{i-1},\tilde{I}_i}(q_{i-1}^\star(\tilde{\mv{I}}))\|+\|{\mv{u}}_F-{\mv{u}}_{\tilde{I}_{\tilde{N}-1},\tilde{I}_{\tilde{N}}}(q_{\tilde{N}-1}^\star(\tilde{\mv{I}}))\|\nonumber\\
\overset{(F_1)}{\leq} & \|{\mv{u}}_0-{\mv{u}}_{{I}_1^\star,{I}_2^\star}(\hat{q}_1({\mv{I}}^\star))\|+\sum_{i=2}^{N^\star-1}\|{\mv{u}}_{{I}_{i}^\star,{I}^\star_{i+1}}(\hat{q}_{i}({\mv{I}}^\star))-{\mv{u}}_{{I}^\star_{i-1},{I}^\star_i}(\hat{q}_{i-1}({\mv{I}}^\star))\|\nonumber\\
&+\|{\mv{u}}_F-{\mv{u}}_{{I}^\star_{{N}^\star-1},{I}^\star_{{N}^\star}}(\hat{q}_{{N}^\star-1}({\mv{I}}^\star))\|,
\end{align}
where $(F_1)$ holds since $({\mv{I}}^\star,\hat{\mv{q}}({\mv{I}}^\star))$ is generally a suboptimal solution to Problem (P3-II) that achieves no smaller objective value compared to the optimal solution $(\tilde{\mv{I}},{\mv{q}}^\star(\tilde{\mv{I}}))$.

Moreover, we have
\begin{align}
&\|{\mv{u}}_0-{\mv{u}}_{I_1^\star,I_2^\star}(\hat{q}_1({\mv{I}}^\star))\|=\|({\mv{u}}_0-{{\mv{u}}^1}^\star)+({{\mv{u}}^1}^\star-{\mv{u}}_{I_1^\star,I_2^\star}(\hat{q}_1({\mv{I}}^\star))\|\nonumber\\
\overset{(F_2)}{\leq} &\|{\mv{u}}_0-{{\mv{u}}^1}^\star\|+\|{{\mv{u}}^1}^\star-{\mv{u}}_{I_1^\star,I_2^\star}(\hat{q}_1({\mv{I}}^\star))\|\nonumber\\
\overset{(F_3)}{\leq}&\|{\mv{u}}_0-{{\mv{u}}^1}^\star\|+2\bar{d}\sin\left(\frac{\theta_{I_{1}^\star,I_2^\star}}{4(Q-1)}\right)\overset{(F_4)}{\leq}
\|{\mv{u}}_0-{{\mv{u}}^1}^\star\|+2\bar{d}\sin\left(\frac{\pi}{4(Q-1)}\right),
\end{align}
where $(F_2)$ holds due to the triangle inequality; $(F_3)$ can be derived based on our applied quantization method; and $(F_4)$ holds since $\theta_{I_i,I_{i+1}}\leq \pi$ holds for any $I_i\neq I_{i+1}$. Similarly, we have
\begin{align}
&\|{\mv{u}}_F-{\mv{u}}_{I_{N^\star-1}^\star,I_{N^\star}^\star}(\hat{q}_{N^\star-1}({\mv{I}}^\star))\|\leq
\|{\mv{u}}_F-{{\mv{u}}^{N^\star-1}}^{\star}\|+2\bar{d}\sin\left(\frac{\pi}{4(Q-1)}\right),\\
& \|{\mv{u}}_{{I}_{i}^\star,{I}^\star_{i+1}}(\hat{q}_{i}({\mv{I}}^\star))-{\mv{u}}_{{I}^\star_{i-1},{I}^\star_i}(\hat{q}_{i-1}({\mv{I}}^\star))\|\leq
\|{{\mv{u}}^i}^\star-{{\mv{u}}^{i-1}}^\star\|+4\bar{d}\sin\left(\frac{\pi}{4(Q-1)}\right).
\end{align}
Therefore, we have the following bounds for $\tilde{s}^{\mathrm{II}}-s^\star$:
\begin{align}
0\leq\tilde{s}^{\mathrm{II}}-s^\star&\leq 2\bar{d}\sin\left(\frac{\pi}{4(Q-1)}\right)+(N^\star-2)\times4\bar{d}\sin\left(\frac{\pi}{4(Q-1)}\right)+2\bar{d}\sin\left(\frac{\pi}{4(Q-1)}\right)\nonumber\\
&\overset{(F_5)}{\leq} 4(M-1)\bar{d}\sin\left(\frac{\pi}{4(Q-1)}\right),
\end{align}
where $(F_5)$ holds since $N^\star\leq M$ due to the non-repeated GBS-UAV association property given in Proposition \ref{lemma_I}.

The proof of Proposition \ref{prop_complexity} is thus completed.

\bibliographystyle{IEEEtran}
\bibliography{Ref_Main}
\end{document}